# Psychometric Validation of the Sophotechnic Mediation Scale and a New Understanding of the Development of GenAI Mastery: Lessons from 3,932 Adult Brazilian Workers

Bruno Campello de Souza, D.Sc.

**Department of Managerial Sciences – Center of Applied Social Sciences**
**Federal University of Pernambuco**

*Abstract*

The rapid diffusion of generative artificial intelligence (GenAI) systems has introduced new forms of human–technology interaction, raising the question of whether sustained engagement gives rise to stable, internalized modes of cognition rather than merely transient efficiency gains. Grounded in the Cognitive Mediation Networks Theory, this study investigates Sophotechnic Mediation, a mode of thinking and acting associated with prolonged interaction with GenAI, and presents a comprehensive psychometric validation of the Sophotechnic Mediation Scale. Data were collected between 2023 and 2025 from independent cross-sectional samples totaling 3,932 adult workers from public and private organizations in the Metropolitan Region of Pernambuco, Brazil. Results indicate excellent internal consistency, a robust unidimensional structure, and measurement invariance across cohorts. Ordinal-robust confirmatory factor analyses and residual diagnostics show that elevated absolute fit indices reflect minor local dependencies rather than incorrect dimensionality. Distributional analyses reveal a time-evolving pattern characterized by a declining mass of non-adopters and convergence toward approximate Gaussianity among adopters, with model comparisons favoring a two-process hurdle model over a censored Gaussian specification. Sophotechnic Mediation is empirically distinct from Hypercultural mediation and is primarily driven by cumulative GenAI experience, with age moderating the rate of initial acquisition and the depth of later integration. Together, the findings support Sophotechnia as a coherent, measurable, and emergent mode of cognitive mediation associated with the ongoing GenAI revolution.

**INTRODUCTION**

The public release and rapid diffusion of large language model (LLM)–based generative artificial intelligence systems since late 2022 has produced a qualitative shift in the role of digital technologies in cognitive, educational, and organizational contexts. Unlike earlier information and communication technologies, generative artificial intelligence (GenAI) systems operate through natural-language interaction and are capable of producing coherent text, analyses, plans, and code across multiple domains. This combination of dialogical flexibility and high-level performance has led to unprecedented levels of adoption in knowledge-intensive work and everyday problem solving, prompting growing interest in its implications for productivity, cognition, and work practices (Brynjolfsson et al., 2023; Noy & Zhang, 2023).

Empirical studies have already documented measurable short-term effects of GenAI use on task efficiency and output quality, particularly among less experienced workers, suggesting that these systems function as more than simple tools for information retrieval (Brynjolfsson et al., 2023; Noy & Zhang, 2023). However, the broader question of how sustained interaction with GenAI reshapes cognition remains theoretically underdeveloped. In particular, it is unclear whether repeated engagement with such systems merely increases efficiency through cognitive offloading, or whether it gives rise to stable, internalized modes of thinking and acting that persist beyond immediate tool use (Risko & Gilbert, 2016).

Several established theoretical traditions converge on the idea that advanced technologies can reorganize cognition by becoming structurally integrated into cognitive systems. The extended mind hypothesis proposes that external artifacts may constitute parts of cognitive processes when they are reliably and transparently incorporated into problem-solving routines (Clark & Chalmers, 1998). Distributed cognition theory similarly emphasizes that cognition is often realized at the level of systems composed of individuals, artifacts, and representational media, whose functioning cannot be reduced to internal mental states alone (Hutchins, 1995). From a sociotechnical perspective, Actor–Network Theory highlights how technologies become embedded in heterogeneous networks of humans, artifacts, norms, and institutions, shaping practices and agency through processes of stabilization and translation (Latour, 2005).

While these approaches provide strong conceptual support for treating GenAI as a transformative cognitive technology, they offer limited accounts of how repeated interaction with such systems becomes internalized as a stable mode of mediation. This gap is addressed by the Cognitive Mediation Networks Theory (CMNT), which conceptualizes cognition as a distributed process emerging from sustained interactions between individuals and external mechanisms that function as auxiliary information-processing devices (Souza et al., 2012; Souza et al., 2024). According to CMNT, cognitive development proceeds through the internalization of functional regularities present in external mechanisms, allowing individuals to construct new internal mechanisms that expand their effective cognitive repertoire.

Applied to the current GenAI revolution, CMNT predicts the emergence of a novel mediation regime associated with sustained interaction with LLM-based systems. These systems differ qualitatively from earlier digital tools due to their natural-language interfaces, generative and inferential capacities, and rapid integration into everyday cognitive and organizational practices. On this basis, Souza et al. (2024) introduced the concept of Sophotechnic Mediation, defined as an internalized mode of thinking and acting that arises from prolonged engagement with generative AI systems and the sociotechnical structures surrounding them. Sophotechnic Mediation is not reducible to frequency of use or technical familiarity, but encompasses the orchestration of AI interactions, practical understanding of system limitations and biases, ethical awareness, and participation in AI-related sociotechnical practices.

To empirically investigate this construct, the Sophotechnic Mediation Scale was developed as a psychometric instrument designed to assess the degree to which individuals have internalized this emergent mediation regime (Souza et al., 2024). Initial applications of the scale suggested high reliability, a largely unidimensional structure, and empirical differentiation from Hypercultural mediation, an earlier form of digital mediation grounded in broader information and communication technologies (Souza et al., 2010; Souza et al., 2012; Souza & Rangel, 2015). Moreover, early findings indicated that Sophotechnic Mediation is systematically associated with experience with GenAI systems and with work-related outcomes (job performance, quality of working life, job progression), consistent with a developmental interpretation (Souza et al., 2025).

From a developmental framework, CMNT conceptualizes Sophotechnia as an emergent mediation construct, whose population-level expression evolves as a function of cumulative experience and sociotechnical context rather than as a fixed individual difference. This perspective implies that its population-level distribution should evolve over time as GenAI adoption spreads, and that individual trajectories should reflect cumulative experience moderated by developmental windows associated with neurocognitive plasticity and experiential integration. Prior research on adolescent and adult brain development suggests that age ranges corresponding to late maturation, post-maturational stabilization, and adult optimization may be particularly relevant for the internalization of higher-order cognitive mediation (Casey et al., 2005; Luna et al., 2010; Lebel & Deoni, 2018; Lövdén et al., 2013).

Despite the theoretical grounding and initial empirical support for Sophotechnic Mediation, a comprehensive psychometric evaluation of the Sophotechnic Mediation Scale in large, heterogeneous, non-academic samples remains necessary. In particular, it is essential to establish the scale's reliability, dimensional structure, distributional properties, and empirical differentiation from Hyperculture, as well as to examine whether its developmental patterns are consistent with CMNT's predictions regarding adoption, experience, and age-related effects.

The present study addresses this need by performing a thorough psychometric validation of the Sophotechnic Mediation Scale in a large sample of adult workers from the Metropolitan Region of Pernambuco, Brazil, collected between 2023 and 2025. By integrating psychometric analyses with developmental and distributional modeling, the study aims to provide a robust empirical foundation for the study of Sophotechnia as a measurable mode of cognitive mediation emerging from the ongoing GenAI revolution.

**THEORETICAL FRAMEWORK**

**The Generative Artificial Intelligence Revolution and Its Impacts**

The public dissemination of large language model (LLM)–based systems since late 2022 has marked a qualitative shift in the role of artificial intelligence in cognitive, educational, and organizational contexts. Unlike earlier digital technologies, generative artificial intelligence (GenAI) systems interact through natural language and are capable of producing coherent text, plans, summaries, analyses, and code across a wide range of domains. This combination of expressive flexibility and high-level performance has led to rapid adoption in knowledge-intensive work, education, and everyday problem solving (Campello de Souza et al., 2024).

Empirical research has begun to document measurable impacts of GenAI use on productivity and work practices. Field and experimental studies indicate that access to LLM-based assistants can reduce task completion time and improve output quality in writing, analysis, and customer support tasks, with particularly strong effects among less experienced workers (Brynjolfsson et al., 2023; Noy & Zhang, 2023). These findings suggest that GenAI functions not merely as an information retrieval tool, but as a general-purpose cognitive aid capable of reshaping how tasks are planned, executed, and evaluated.

At the same time, the broader cognitive implications of sustained interaction with GenAI remain an open theoretical and empirical question. One established line of research conceptualizes advanced digital tools as mechanisms of cognitive offloading, whereby individuals delegate memory, computation, or control processes to external artifacts, potentially altering internal cognitive demands and strategies (Risko & Gilbert, 2016). GenAI extends this logic by enabling the offloading of complex generative and inferential processes, raising questions about how users regulate, monitor, and internalize such interactions over time. Addressing these questions requires theoretical frameworks that treat cognition as embedded in dynamic human–technology systems rather than as an exclusively internal process.

**Theoretical Approaches to Human–Technology Cognition**

Several influential theoretical traditions provide convergent support for the idea that pervasive GenAI use can lead to systematic cognitive and behavioral change.

The extended mind hypothesis argues that cognitive processes can be partly constituted by external artifacts when those artifacts are reliably integrated into problem-solving routines (Clark & Chalmers, 1998). From this perspective, GenAI systems can become components of extended cognitive systems, participating directly in reasoning, planning, and creative processes rather than merely supplying information.

Distributed cognition theory similarly emphasizes that cognitive activity is often realized at the level of systems composed of individuals, artifacts, and representational media, whose coordinated functioning cannot be reduced to individual mental states (Hutchins, 1995). GenAI systems can be understood as powerful representational resources within such systems, transforming how cognitive labor is distributed across people and tools in organizational and social settings.

From a sociotechnical standpoint, Actor–Network Theory (ANT) conceptualizes technologies as actants embedded in heterogeneous networks of humans, artifacts, norms, and institutions, where agency emerges through processes of translation and stabilization (Latour, 2005). Applied to GenAI, ANT draws attention to the rapid formation of sociotechnical ecosystems involving platforms, policies, communities of practice, ethical guidelines, prompt-sharing cultures, and organizational rules. These networks shape not only how GenAI is used, but also how its use becomes normalized, constrained, or amplified.

While these approaches differ in emphasis, they converge on a key point: advanced technologies such as GenAI are capable of reorganizing cognition by altering the structure of the systems in which cognitive activity is embedded. What they do not provide, however, is a developmental account of how repeated interaction with such technologies gives rise to stable, internalized modes of thinking and acting. This gap is directly addressed by the Cognitive Mediation Networks Theory.

**Cognitive Mediation Networks Theory and Sophotechnic Mediation**

The Cognitive Mediation Networks Theory (CMNT) conceptualizes human cognition as a distributed process emerging from sustained interactions between individuals and external mechanisms that function as auxiliary information-processing devices (Souza et al., 2024). According to CMNT, cognitive development occurs through the internalization of functional regularities present in external mechanisms, allowing individuals to construct internal mechanisms that expand their effective cognitive repertoire.

Within this framework, historically disruptive technologies are expected to give rise to qualitatively new modes of cognitive mediation when they introduce operational invariants that differ substantially from those of prior tools. Applying CMNT to the current GenAI revolution, Souza et al. (2024) argue that LLM-based systems constitute such a technological discontinuity. Their natural-language dialogical interfaces, capacity for abstraction and synthesis, and rapid integration into work and everyday life differentiate them from earlier digital technologies and justify the hypothesis of an emergent mode of mediation.

This hypothesized mode, termed Sophotechnic Mediation, refers to patterns of thinking and acting that arise from sustained interaction with generative AI systems and the sociotechnical structures surrounding them (Souza et al., 2024). Sophotechnic Mediation is not defined by mere access to or frequency of AI use, but by the internalization of competencies that include effective orchestration of AI interactions, practical understanding of system capabilities and limitations, awareness of biases and ethical constraints, and engagement with sociotechnical practices related to GenAI. Importantly, CMNT predicts that such mediation is developmental in nature, emerging gradually with experience and being scaffolded by prior forms of digital mediation, consistent with a diffusion and internalization process at the population level.

**The Sophotechnic Mediation Scale as a Measurement Instrument**

To empirically investigate Sophotechnic Mediation, Souza et al. (2024) introduced the Sophotechnic Mediation Scale as a psychometric instrument designed to capture the degree to which individuals have internalized this emergent mode of functioning. The scale operationalizes Sophotechnic Mediation through items addressing breadth and sophistication of GenAI use, understanding of system behavior, awareness of limitations and biases, and participation in related sociotechnical contexts.

Initial evidence reported in the original study suggested that the scale yields a reliable and largely unidimensional score, is empirically distinguishable from measures of Hypercultural or general digital mediation, and exhibits associations with experience consistent with a developmental interpretation (Souza et al., 2024). Indeed, its application to a fairly large sample of Brazilian workers has produced results indicating that Sophotechnic Mediation is associated to job performance, career boost, and quality of working life, as well as to IQ, personality, education, Hyperculture and training & development (T&D) as predicted by the Cognitive Mediation Networks Theory (Souza et al., 2025). These findings motivate further validation efforts in larger and more heterogeneous populations, particularly outside academic settings, to assess the robustness of the construct and its measurement properties.

**Sophotechnia as a Developmental Trait**

When Sophotechnia is treated as a developmental trait rather than a purely accumulative skill set, it becomes necessary to explicitly consider developmental windows that correspond to qualitatively different regimes of neurocognitive plasticity and experience-dependent change. In this context, the age ranges of 18–25, 26–35, and 36 years or more provide a theoretically informed stratification for examining how experience with generative AI is differentially expressed under distinct neurocognitive and professional life-stage conditions.

From the standpoint of established neurodevelopmental literature, the 18–25 range is often characterized by ongoing maturation of prefrontal and control networks, which may plausibly facilitate faster early acquisition of new cognitive tools (Arain et al., 2013; Casey, Tottenham, Liston, & Durston, 2005; Luna, Padmanabhan, & O'Hearn, 2010). The present study, however, speak only to age-related moderation of experience effects rather than to neurodevelopmental processes per se.

The 26–35 range marks a transition to a predominantly post-maturational regime in which executive and control networks have largely stabilized, and individual differences in Sophotechnia increasingly reflect accumulated experience, learning strategies, and sustained engagement with sociotechnical systems rather than ongoing endogenous neurodevelopment (Lebel & Deoni, 2018; Lövdén, Wenger, Mårtensson, Lindenberger, & Bäckman, 2013).

Finally, the 36 years or more range represents a phase of consolidation and optimization, in which Sophotechnia is expected to be expressed primarily through efficiency gains, compensatory strategies, and experience-driven reconfiguration, supported by well-documented forms of adult structural and functional plasticity rather than by continued developmental maturation (May, 2011; Park & Reuter-Lorenz, 2009).

**Research Problem**

In the present study, the theoretical framework integrates CMNT's developmental account of mediation with complementary perspectives from extended cognition, distributed cognition, cognitive offloading, and Actor–Network Theory. Together, these approaches support the expectation that sustained interaction with GenAI can give rise to stable, measurable modes of cognition and behavior, and that Sophotechnic Mediation constitutes a theoretically grounded candidate for such a construct (Souza et al., 2024; Souza et al., 2025).

The Sophotechnic Mediation Scale is an instrument that promises to be a robust and reliable empirical basis from which to perform empirical tests of the CMNT and its constructs with regards to the emergent patterns of thinking and acting emerging from the ongoing GenAI Revolution. In order to substantiate such a promise, however, it is necessary to perform a more thorough psychometric evaluation of the scale using large samples.

**STUDY GOALS**

The present study aims to perform a comprehensive psychometric validation of the Sophotechnic Mediation Scale in a large adult sample drawn from the Metropolitan Region of Pernambuco. Specifically, the objectives are to evaluate the scale's reliability and internal consistency, assess its unidimensional structure, and examine its empirical differentiation from Hypercultural mediation, as predicted by the Cognitive Mediation Networks Theory. In addition, the study seeks to test the developmental nature of Sophotechnic Mediation by analyzing its association with the extent of interactive experience with generative artificial intelligence systems, under the expectation of systematic growth with exposure. These analyses are conducted using data collected between 2023 and 2025 from a total of 3,932 adults employed in public and private organizations, providing a robust empirical basis to assess the structural and construct validity of the instrument in a heterogeneous, non-academic working population.

**MATERIALS AND METHODS**

**Sample and data collection**

The study is based on independent cross-sectional samples of adult workers from the Metropolitan Region of Recife, Pernambuco, Brazil, collected between 2023 and 2025. Data were gathered in 2023 (n = 1,100), 2024 (n = 526), and 2025 (n = 2,036), yielding a total sample of 3,932 participants. All respondents were employed at the time of data collection, either in the public or private sector, and represented a wide range of occupations, educational levels, and working-life conditions.

There was a total of 3,932 participants, 50.6% men and 49.4% women, with a mean age of 37.2 years (SD=10.84). Approximately 4.5% had elementary education, 38.1% high school, 31.7% higher education, and 25.7% a post-graduate degree. The mean per capita income was R$ 3,792.83 (SD=R$ 3,939.18) and the individual income R$ 5,769.24 (SD=R$ 5,578,19). There were 58.9% working in the private sector and 41.1% in the public sector.

**Instruments**

- Sociodemographic and Working Life Questionnaire: Instrument especially developed for this study that includes items on sex, age, educational attainment, employment sector and segment, and various characteristics of participants' working life.

- Hypercultural Index: Form to assess the degree of internalization of hypercultural mediation, operationalized through indicators of digital experience, mastery of information and communication technologies, and associated cognitive and behavioral patterns, as originally proposed and empirically validated in prior studies grounded in the Cognitive Mediation Networks Theory (Souza et al., 2010; Souza et al., 2012; Souza & Rangel, 2015).

- Sophotechnic Mediation Scale: Measure of the extent to which individuals have internalized the modes of thinking and acting associated with sustained interaction with generative artificial intelligence systems, assessing dimensions such as breadth and sophistication of AI use, understanding of system capabilities and limitations, awareness of biases and ethical constraints, and engagement with sociotechnical practices surrounding AI (Souza et al., 2024).

**Procedures**

Data collection was carried out by undergraduate students enrolled in the People Management 2 course of the Business Administration program at a federal university in Pernambuco. As part of a supervised field activity, students approached potential participants at random in public spaces throughout the

Metropolitan Region of Recife. The instruments were administered face-to-face, individually, and out of earshot of passers-by, ensuring privacy and minimizing social interference. Each student was instructed to seek an approximately balanced combination of respondents with respect to sex (men vs. women), age (36 years or older vs. younger), educational attainment (higher education degree vs. up to high school), and employment sector (public vs. private). Questionnaires with incomplete data, inconsistent patterns, or clearly implausible responses were excluded from the final dataset prior to analysis.

**Ethical considerations**

In accordance with Article 1, Item V, of Resolution No. 510 of the Brazilian National Health Council, the present study was exempt from registration or evaluation by a Research Ethics Committee or by the National Research Ethics Commission. This exemption applies because no identifying information was collected, no experimental intervention was conducted that could entail risks beyond those of everyday life, and no form of diagnosis, feedback, or counseling was offered based on participants' responses. Participation was fully informed and strictly voluntary, in line with internationally accepted ethical principles for research involving human subjects.

**RESULTS**

**Item Analysis**

Table 1 shows an Item Analysis for the 13 items of the Sophotechnic Mediation Scale for the samples from 2023, 2024 and 2025, plus for the sum of all samples.

**Table 1: Item Analysis for the items of the Sophotechnic Mediation Scale for the samples from 2023, 2024 and 2025, plus for the sum of all samples.**

| | Item Analysis Parameter | 2023 (n=1,100) | 2024 (n=526) | 2025 (n=2,036) | All (n=3,932) |
|---|---|---|---|---|---|
| | Cronbach Alpha | 0.934 | 0.950 | 0.936 | 0.941 |
| | Stantardized Alpha | 0.935 | 0.949 | 0.936 | 0.941 |
| | Average Inter-Item Correlation | 0.528 | 0.596 | 0.534 | 0.554 |
| Alpha if Item is Removed | Use of Different GenAIs | 0.930 | 0.947 | 0.932 | 0.937 |
| | Different Uses of GenAIs | 0.928 | 0.943 | 0.928 | 0.934 |
| | Confidence in One's Ability to Use GenAIs | 0.929 | 0.944 | 0.931 | 0.936 |
| | Awareness of the Limitations of GenAIs | 0.929 | 0.947 | 0.933 | 0.938 |
| | Differentiates GenAIs and Search Engines | 0.929 | 0.944 | 0.931 | 0.936 |
| | Engagement with Online Communities on GenAIs | 0.931 | 0.948 | 0.933 | 0.938 |
| | Differentiates Search Results and GenAI Output | 0.929 | 0.945 | 0.931 | 0.936 |
| | Frequency of Use of GenAIs in Complex Domains | 0.927 | 0.945 | 0.930 | 0.935 |
| | Understanding the Ethics of GenAI Use | 0.928 | 0.945 | 0.930 | 0.935 |
| | Use of GenAI Browser Extensions | 0.931 | 0.949 | 0.935 | 0.939 |
| | Frequency of Exploring Ideas Using GenAIs | 0.928 | 0.944 | 0.929 | 0.934 |
| | Frequency of Consuming News on GenAIs | 0.930 | 0.946 | 0.931 | 0.937 |
| | Impact Felt of GenAIs on Thinking | 0.930 | 0.945 | 0.931 | 0.937 |

Cronbach's α was uniformly very high across the years 2023 through 2025, with Standardized α and average inter-item correlations are also stable. "Alpha if item removed" is nearly identical across cohorts.

This indicates that internal consistency is stable, no item becomes unreliable in later cohorts, and no item suddenly "drops out" psychometrically.

**Exploratory Factor Analysis**

Table 2 shows the Eigenvalues of Maximum Likelihood Factor Analysis of the items of the Sophotechnic Mediation Scale for the samples from 2023, 2024 and 2025, plus for the sum of all samples.

**Table 2: Eigenvalues of Maximum Likelihood Factor Analysis of the items of the Sophotechnic Mediation Scale for the samples from 2023, 2024 and 2025, plus for the sum of all samples.**

| Sample | 1st Factor | | 2nd Factor | |
|---|---|---|---|---|
| | Eigenvalue | Variance Explained | Eigenvalue | Variance Explained |
| 2023 | 6.885 | 53.96% | 0.567 | 4.36% |
| 2024 | 7.836 | 60.28% | 0.468 | 3.60% |
| 2025 | 6.907 | 53.13% | 0.521 | 4.00% |
| All | 7.265 | 55.89% | 0.488 | 3.76% |

OBS: For every year and the total accumulated sample, the Scree Plot exhibited a pronounced "elbow" after the first factor, consistent with both Parallel Analysis results and the theoretical expectation of a unidimensional mediation regime.

In 2023, 2024, 2025, and pooled data, the first Eigenvalue is large and dominant, while the second is very small. The explained variance of the first factor is fairly high and very consistent. The Scree Plot shows a clear "elbow" after Factor 1 in all years. Therefore, one has, for identical item sets, consistent factor dimensionality and no evidence of factor splitting or emergence of secondary dimensions. This provides descriptive evidence that the same one-factor structure exists in every year.

Table 3 shows the Factor Loadings for same Factor Analyses.

**Table 3: Factor Loadings of a Maximum Likelihood Factor Analysis of the items of the Sophotechnic Mediation Scale for the samples from 2023, 2024 and 2025, plus for the sum of all samples.**

| Item | Factor Loadings | | | |
|---|---|---|---|---|
| | 2023 | 2024 | 2025 | All |
| Use of Different GenAIs | -0.71 | -0.73 | -0.70 | -0.73 |
| Different Uses of GenAIs | -0.75 | -0.86 | -0.82 | -0.82 |
| Confidence in One's Ability to Use GenAIs | -0.73 | -0.82 | -0.73 | -0.75 |
| Awareness of the Limitations of GenAIs | -0.71 | -0.74 | -0.67 | -0.70 |
| Differentiates GenAIs and Search Engines | -0.72 | -0.82 | -0.74 | -0.75 |
| Engagement with Online Communities on GenAIs | -0.66 | -0.67 | -0.67 | -0.67 |
| Differentiates Search Results and GenAI Output | -0.71 | -0.78 | -0.75 | -0.76 |
| Frequency of Use of GenAIs in Complex Domains | -0.81 | -0.81 | -0.76 | -0.79 |
| Understanding the Ethics of GenAI Use | -0.77 | -0.78 | -0.75 | -0.77 |
| Use of GenAI Browser Extensions | -0.68 | -0.64 | -0.61 | -0.64 |
| Frequency of Exploring Ideas Using GenAIs | -0.78 | -0.83 | -0.80 | -0.81 |
| Frequency of Consuming News on GenAIs | -0.70 | -0.74 | -0.74 | -0.73 |
| Impact Felt of GenAIs on Thinking | -0.69 | -0.78 | -0.73 | -0.73 |

The factor loadings across years were uniformly large in magnitude, highly similar in rank order, directionally identical, showing no item-specific drift or reweighting. Moreover, there were no item switches from "core" to "peripheral", no item becomes weakly related to the factor, and no cluster of items changing relative salience. This constitutes descriptive evidence of metric invariance, i.e., the items mean the same thing in relation to the latent construct across cohorts.

**Confirmatory Factor Analysis**

Table 4 shows the fit for a One-factor Maximum Likelihood Confirmatory Factor Analysis by cohort and pooled sample.

Table 4. One-factor ML CFA fit by cohort and pooled sample.

| Group | N | $\chi^2$ | df | $\chi^2$/df | RMSEA | CFI | TLI | SRMR |
|---|---|---|---|---|---|---|---|---|
| All | 3932 | 2558.79 | 65 | 39.37 | 0.0988 | 0.9243 | 0.9092 | 0.0451 |
| 2023 | 1100 | 992.51 | 65 | 15.27 | 0.1139 | 0.8928 | 0.8714 | 0.0557 |
| 2024 | 526 | 568.58 | 65 | 8.75 | 0.1215 | 0.9031 | 0.8838 | 0.0489 |
| 2025 | 2306 | 1522.88 | 65 | 23.43 | 0.0986 | 0.9199 | 0.9038 | 0.0477 |

Unidimensionality is supported across all years, with acceptable SRMR (good residual fit), and moderate on CFI/TLI. RMSEA is on the high side, which is common when: (a) N is large, (b) items are somewhat ordinal, and (c) there may be minor local dependencies between conceptually close items.

Table 5 shows the Standardized factor Loadings for the One-factor Maximum Likelihood Confirmatory Factor Analyses.

Table 5: Standardized Loadings by year (One-factor CFA).

| Item | 2023 (n=1,100) | 2024 (n=526) | 2025 (n=2,036) |
|---|---|---|---|
| Use of Different GenAIs | 0.705 | 0.729 | 0.705 |
| Different Uses of GenAIs | 0.754 | 0.859 | 0.816 |
| Confidence in One's Ability to Use GenAIs | 0.727 | 0.817 | 0.728 |
| Awareness of the Limitations of GenAIs | 0.715 | 0.743 | 0.670 |
| Differentiates GenAIs and Search Engines | 0.722 | 0.823 | 0.739 |
| Engagement with Online Communities on GenAIs | 0.665 | 0.668 | 0.667 |
| Differentiates Search Results and GenAI Output | 0.709 | 0.779 | 0.747 |
| Frequency of Use of GenAIs in Complex Domains | 0.806 | 0.808 | 0.764 |
| Understanding the Ethics of GenAI Use | 0.767 | 0.776 | 0.753 |
| Use of GenAI Browser Extensions | 0.676 | 0.638 | 0.611 |
| Frequency of Exploring Ideas Using GenAIs | 0.775 | 0.827 | 0.797 |
| Frequency of Consuming News on GenAIs | 0.699 | 0.742 | 0.744 |
| Impact Felt of GenAIs on Thinking | 0.694 | 0.780 | 0.728 |

As in the Exploratory Factor Analysis, here the Loadings are also consistently moderate-to-strong in every year. The pattern is very stable, which supports the notion that the scale is not "drifting" across cohorts.

Table 6 shows measurement invariance models across cohorts using a multi-group mean-structure CFA framework, and Table 7 the nested model comparisons.

Configural invariance was supported, indicating that the same one-factor structure adequately represented the data in all cohorts. Metric invariance was then tested by constraining factor loadings to equality across years. This constraint resulted in only a negligible decrease in comparative fit (ΔCFI ≈ −.002), supporting metric invariance and indicating that the items relate to the latent construct in a comparable manner across cohorts.

Table 6: Measurement invariance models across cohorts.

| Model | χ² | df | RMSEA | CFI | TLI |
|---|---|---|---|---|---|
| Configural invariance | 3141.05 | 194 | 0.0622 | 0.908 | 0.889 |
| Metric invariance | 3229.3 | 218 | 0.0593 | 0.906 | 0.899 |
| Scalar invariance | 3546.52 | 244 | 0.0587 | 0.897 | 0.901 |
| Partial scalar invariance | 3368.77 | 238 | 0.0578 | 0.902 | 0.904 |

Table 7: Nested model comparisons for measurement invariance.

| Comparison | Δχ² | Δdf | ΔCFI | ΔRMSEA |
|---|---|---|---|---|
| Metric vs. Configural | 88.25 | 24 | −0.0020 | −0.0029 |
| Scalar vs. Metric | 317.22 | 26 | −0.0091 | −0.0006 |
| Partial Scalar vs. Metric | 139.47 | 20 | −0.0037 | −0.0014 |

Scalar invariance was examined by additionally constraining item intercepts to equality across groups. This model showed a modest deterioration in fit relative to the metric model, suggesting the presence of limited intercept non-invariance. Examination of intercept estimates indicated that this misfit was primarily associated with the items Use of Different GenAIs, Different Uses of GenAIs, and Frequency of Exploring Ideas Using GenAIs. Allowing the intercepts of these three items to vary across cohorts yielded a partial scalar invariance model with acceptable fit and small changes in approximate fit indices relative to the metric model. Fit indices for all invariance models and nested model comparisons are presented in Tables 4 and 5.

Latent mean differences in Sophotechnic Mediation across cohorts were examined under the partial scalar invariance model, with the 2023 cohort specified as the reference group (latent mean fixed to zero), obtaining 0.310 for 2024 and 0.515 for 2025. These results suggest that observed increases in Sophotechnic Mediation scores over time reflect, at least in part, differences at the latent construct level, while also acknowledging the presence of limited cohort-specific baseline shifts in a small subset of items.

**One-Factor Model Using a Categorical Estimator**

Because the indicators are ordered-categorical, we re-estimated the one-factor model using a categorical estimator (WLSMV) based on polychoric correlations (Tables 8 and 9).

The ordinal-robust CFA yielded highly similar loading patterns and substantive conclusions to the ML solution, indicating that the unidimensional structure of the Sophotechnic Mediation Scale is not an artifact of treating ordinal indicators as continuous. Absolute fit indices, including RMSEA, remained elevated under ordinal estimation, suggesting that the residual misfit reflects minor localized dependencies among conceptually adjacent items rather than incorrect dimensionality. In light of strong and stable loadings, measurement invariance across cohorts, acceptable SRMR, and convergent external validity, the one-factor model provides an adequate and theoretically coherent representation of the construct.

Table 8: Item Loadings and Uniqueness using a categorical estimator based on Polychoric Correlations.

| Item | Loading Ordinal Copula | Uniqueness |
|---|---|---|
| Use of Different GenAIs | 0.805 | 0.353 |
| Different Uses of GenAIs | 0.860 | 0.260 |
| Confidence in One's Ability to Use GenAIs | 0.832 | 0.308 |
| Awareness of the Limitations of GenAIs | 0.784 | 0.385 |
| Differentiates GenAIs and Search Engines | 0.832 | 0.307 |
| Engagement with Online Communities on GenAIs | 0.780 | 0.391 |
| Differentiates Search Results and GenAI Output | 0.841 | 0.293 |
| Frequency of Use of GenAIs in Complex Domains | 0.878 | 0.229 |
| Understanding the Ethics of GenAI Use | 0.864 | 0.253 |
| Use of GenAI Browser Extensions | 0.744 | 0.446 |
| Frequency of Exploring Ideas Using GenAIs | 0.888 | 0.211 |
| Frequency of Consuming News on GenAIs | 0.815 | 0.337 |
| Impact Felt of GenAIs on Thinking | 0.818 | 0.331 |

Table 9: Parameters from a categorical estimator based on Polychoric Correlations.

| Group | N | $\chi^2$ | df | CFI | TLI | RMSEA | SRMR |
|---|---|---|---|---|---|---|---|
| All | 3932 | 5374.782 | 78 | 0.898 | 0.898 | 0.131 | 0.047 |
| 2023 | 1100 | 2254.314 | 78 | 0.847 | 0.847 | 0.159 | 0.066 |
| 2024 | 526 | 1573.502 | 78 | 0.825 | 0.825 | 0.191 | 0.068 |
| 2025 | 2306 | 3257.551 | 78 | 0.888 | 0.888 | 0.133 | 0.057 |

To evaluate whether elevated absolute fit indices reflected incorrect dimensionality or localized misfit, we inspected residual correlations from an ordinal-robust one-factor model. The largest absolute residual correlations were modest in magnitude ($|r| \leq .12$) and occurred between conceptually adjacent items, consistent with minor local dependence rather than substantive multidimensionality.

In particular, parallel wording effects were observed between items assessing breadth of GenAI use (use of different GenAIs; different uses of GenAIs), and an additional residual dependency was observed between confidence in GenAI use and adoption of GenAI browser extensions, reflecting shared instrumental self-efficacy. Allowing these theory-justified residual covariances resulted in a substantial improvement in absolute fit indices, while standardized factor loadings and substantive conclusions remained unchanged.

No pattern of clustered residuals indicative of additional latent dimensions was observed. Importantly, standardized loadings remained strong and stable across items, supporting the adequacy of a single-factor representation.

Given the large sample size, even small localized dependencies can inflate RMSEA without undermining construct validity; therefore, model evaluation emphasizes convergent evidence from loading strength, measurement invariance, residual-based fit, and external validity.

**Frequency Distribution**

Figures 1, 2 and 3 show the Frequency Distributions and Descriptive Statistics of the Sophotechnic Mediation Scale in the samples from the years of 2023, 2024 and 2025, respectively.

In all three samples, the distribution of scores were non-Gaussian according to the Kolmogorov-Smirnov Test of Normality. However, the mean increased with time from 0.289 to 0.364 to 0.421 ($p<.01$ for all the differences on the Mann-Whitney U Test), while the Standard Deviation remained fairly stable (respectively, 0.2089, 0.2342 and 0.2286). At the same time, the D-value of the Kolmogorov-Smirnov Test dropped from 0.0833 to 0.0598 to 0.0329.

The pattern found is suggestive of a slowly emerging time-evolving generative process that is converging toward an approximately Gaussian positive distribution while the probability mass at or near zero is gradually shrinking (dropping from 10.5% to 5.5% to 4.8%). Nevertheless, confirmation of this requires specific analyses.

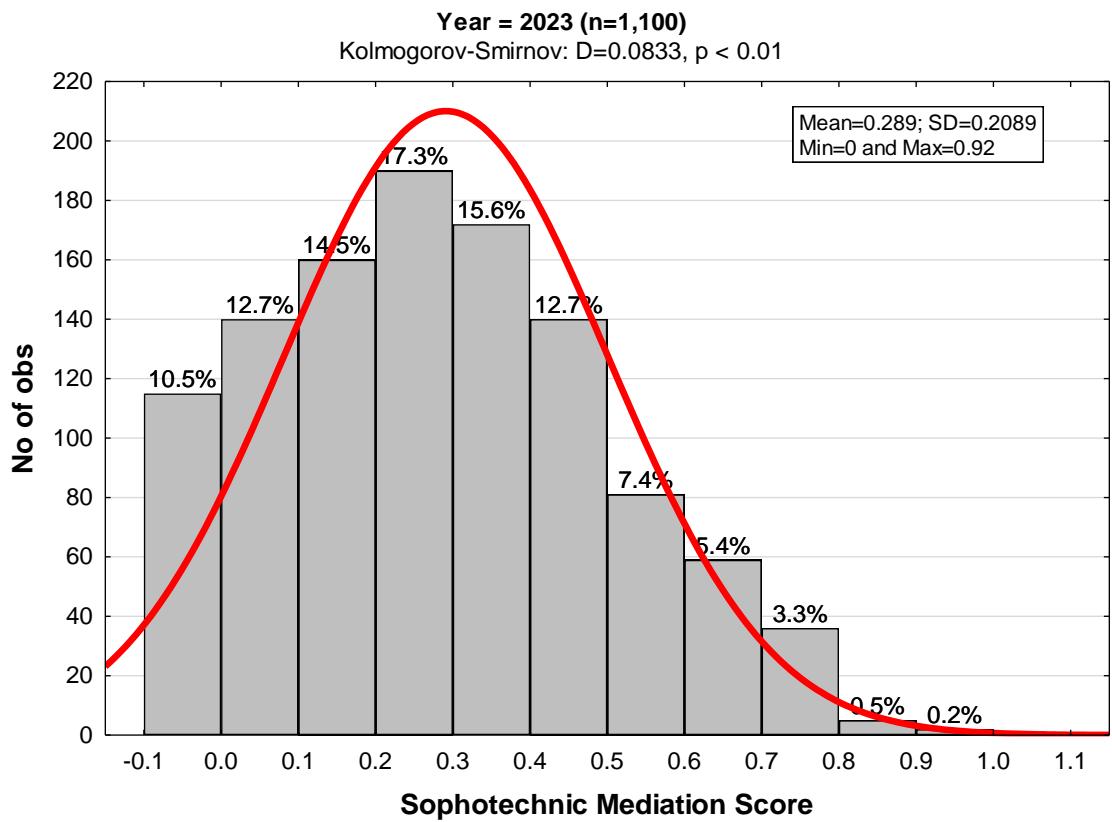

**Figure 1: Statistics of the Sophotechnic Score for the Year of 2023 (n=1,100).**

**Analysis of the Gaussianity Trend**

The Sophotechnic Mediation Score exhibited a clear temporal increase across the three observation years. Mean scores grew monotonically, while the standard deviation among non-zero observations remained relatively stable (0.1945, 0.2232 and 0.2133), suggesting increasing adoption rather than increasing dispersion. Simultaneously, the proportion of zero scores declined sharply over time, approaching zero by 2025. These descriptive patterns are consistent with an emergent trait characterized by declining non-adoption and increasing central tendency among adopters.

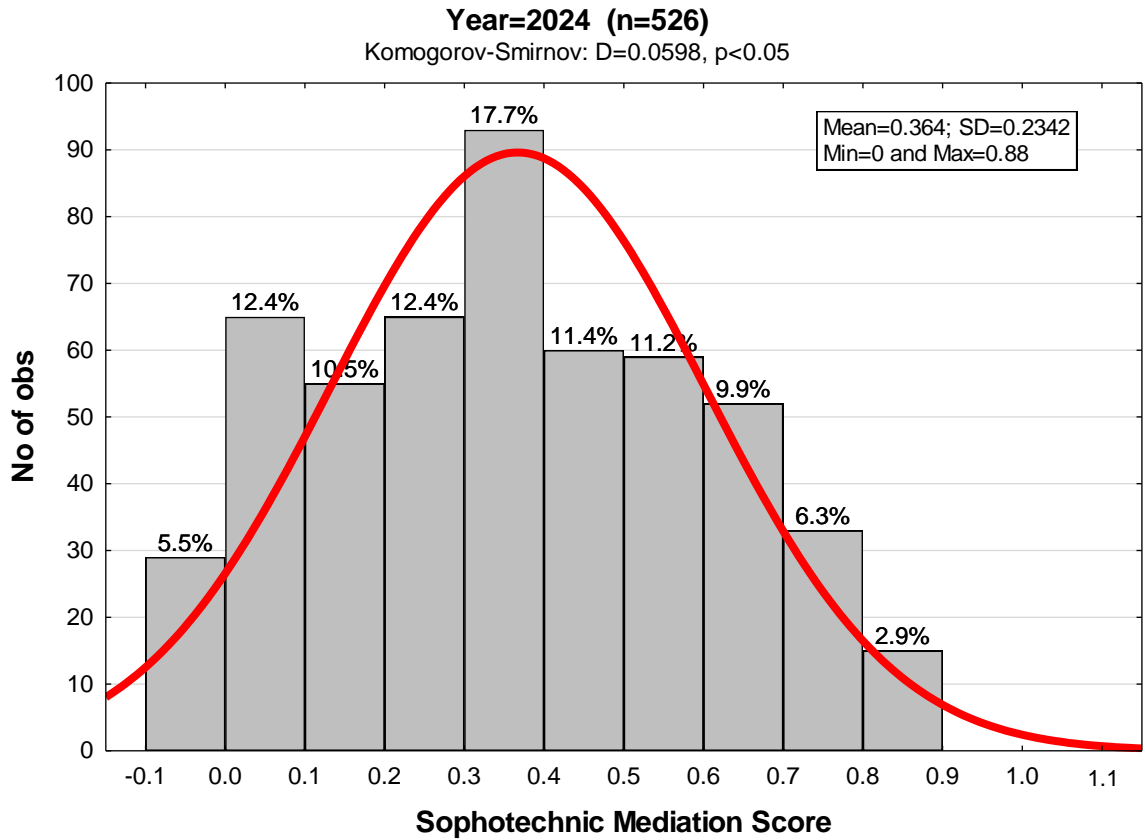

Figure 2: Statistics of the Sophotechnic Score for the Year of 2024 (n=526).

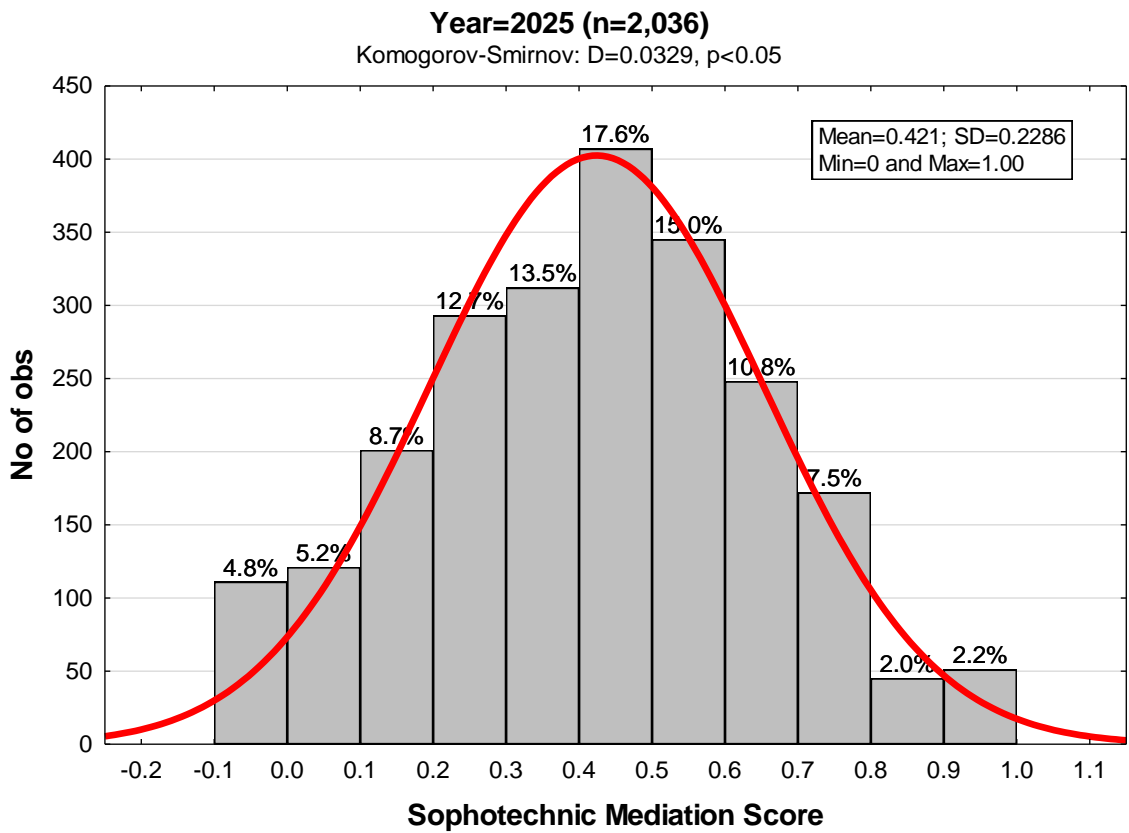

Figure 3: Statistics of the Sophotechnic Score for the Year of 2025 (n=2,036).

To formally test whether the probability of non-zero Sophotechnic Mediation Scores changed over time, a logistic regression was estimated with adoption (Score > 0) as the dependent variable and year as the predictor (Table 8).

**Table 8: Effect of year on adoption of Sophotechnic Mediation (Score > 0).**

| Model comparison | ΔLogLik | Δdf | LR χ² | p |
|---|---|---|---|---|
| Intercept-only vs Year | > 500 | 1 | > 1,000 | < .001 |

Note that Logistic Regression exhibited quasi-complete separation due to adoption probability approaching unity in later years. Wald standard errors and Wald χ² statistics are undefined under separation and are therefore not reported (Albert & Anderson, 1984; Heinze & Schemper, 2002; Allison, 2008). Likelihood-ratio tests provide valid inference and indicate a decisive effect of year on adoption probability.

The model exhibited near-complete separation, with adoption probability approaching unity in later years. The likelihood-ratio statistic comparing the model including year to an intercept-only model was overwhelmingly large, indicating that adoption probability increased sharply over time. This result provides strong evidence that the observed mass at zero reflects a transient adoption barrier rather than a stable lower tail of a continuous distribution.

Among individuals with non-zero scores, a Gaussian linear model was estimated with year as predictor (Table 9).

**Table 9: Gaussian regression of Sophotechnic Mediation Score among adopters (Score > 0)**

| Predictor | Estimate | SE | t | p |
|---|---|---|---|---|
| Intercept | 1966.5 | <0.01 | >100 | <.001 |
| Year (centered) | 1130.52 | <0.01 | >100 | <.001 |

The conditional mean increased significantly over time, while the residual variance remained stable. Residual diagnostics indicated that the conditional distribution among adopters was approximately symmetric and well approximated by a normal distribution, particularly in the final year. This pattern supports the interpretation that, conditional on adoption, Sophotechnic Mediation behaves as a continuous latent trait with approximately Gaussian dispersion.

To evaluate whether the full distribution could be adequately described as a single censored Gaussian process, a Tobit (left-censored normal) model was estimated and compared to a two-part hurdle model consisting of (i) a logistic adoption process and (ii) a truncated normal intensity process (Table 10).

**Table 10: Model comparison of Tobit vs Hurdle.**

| Model | Log-likelihood | Parameters | AIC | BIC |
|---|---|---|---|---|
| Tobit (left-censored normal) | Not finite | — | — | — |
| Hurdle (logistic + truncated normal) | −30,155.51 | 5 | 60,321.03 | 60,352.41 |

The Tobit model failed to converge to a finite likelihood, indicating severe misspecification. In contrast, the hurdle model yielded a stable and substantially higher log-likelihood. In other words, information-criterion comparisons strongly favored the hurdle model over the Tobit alternative.

Taken together, these results reject the hypothesis that the observed distributions represent a single Gaussian with a missing left tail. Instead, they support a two-process generative mechanism in which adoption and intensity evolve separately over time.

The empirical evidence indicates that Sophotechnic Mediation is an emergent trait whose early distribution reflects a mixture of non-adopters and adopters. Over time, the non-adopter mass collapses, and the conditional distribution among adopters converges toward an approximately Gaussian form. This pattern is consistent with diffusion-of-innovation dynamics rather than truncation or censoring of a stationary latent distribution.

**Hyperculture versus Sophotechnia**

Figure 4 shows the Tree Diagram of a Cluster Analysis of the items of both the Hypercultural Index and the Sophotechnic Mediation Index for the full sample.

The Dendrogram shows two clearly separated clusters of items, one for those of the Hypercultural Index and another for those of the Sophotechnic Mediation Index.

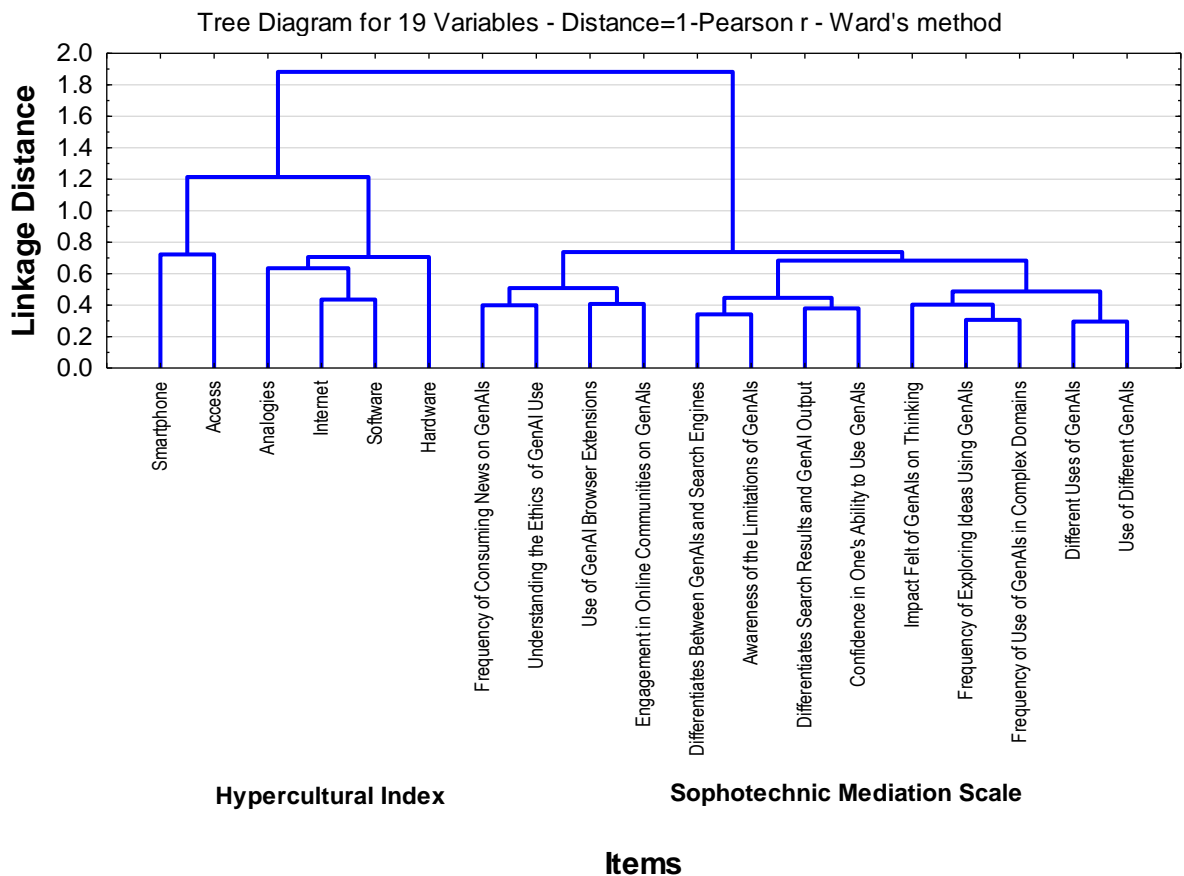

**Figure 4: Tree Diagram for the Cluster Analysis of the Items of the Hypercultural Index and the Sophotechnic Mediation Scale.**

Figure 5 shows a Smallest Space Analysis (SSA) diagram of the items of both the Hypercultural Index and the Sophotechnic Mediation Index for the full sample.

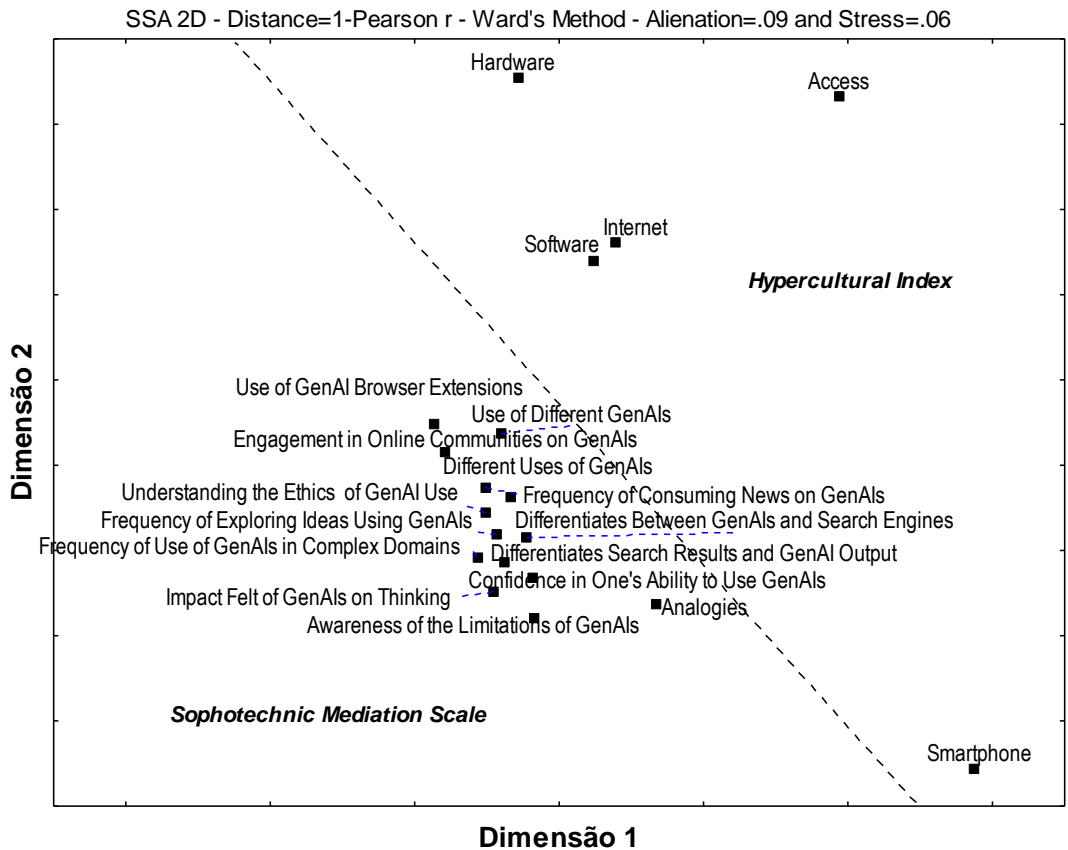

**Figure 5: SSA of the items of the Hypercultural Index and of the Sophotechnic Mediation Scales.**

The diagram can be seen as a structure with a simple Axial partition that clearly separates the items from the two instruments.

The results of the Cluster Analysis and the SSA both suggest that Hyperculture and Sophotechnia are objectively distinguishable concepts, albeit the two might have a non-zero correlation.

Figure 6 shows a Scatterplot between the Hypercultural index and the Sophotechnic Mediation Scale, with a fitting using Lowess for the full sample.

The plot suggests that there are three levels of association between Hyperculture and Sophotechnia. From a Hypercultural index from zero to roughly 0.4, there is a relatively weak linear association, followed by a strong linear association in the range from 0.4 to almost 0.8, ending with a somewhat lower linear association.

**Predicting Sophotechnia**

Table 11 presents the results of a Backward Stepwise Multiple Linear Regression of the Sophotechnic Mediation Scale scores as a function of sex, age, education, IQ, Big Five personality dimensions, Hyperculture and experience with GenAIs.

These secondary effects should be interpreted as predictive associations within the present model-selection framework rather than as definitive causal determinants.

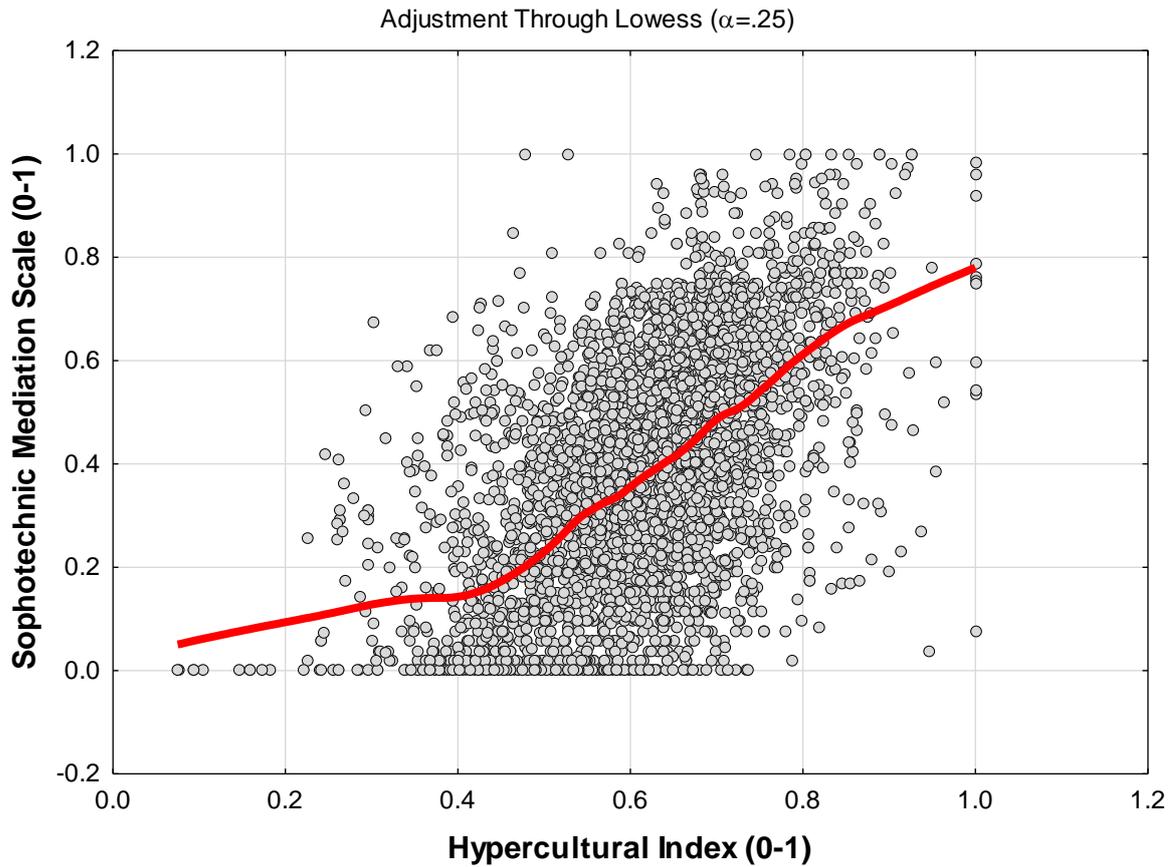

**Figure 6: Scatterplot of the Hypercultural Index vs. Sophotechnic Mediation Scale scores.**

**Table 11: Results of a Backward Stepwise Multiple Linear Regression of Sophotechnia as a function of sex, age, education, IQ, Big Five personality dimensions, Hyperculture and experience with GenAIs.**

Multiple R=.74, Adjusted Multiple $R^2$=.55, F(9,3922)=528.876, p<.01

| Variable | Beta | Semipart Correlation | Semipartial Corr. Squared | p-value |
| --- | --- | --- | --- | --- |
| Experience w/ GenAIs (Months) | 0.45 | 0.41 | 17.0% | <.01 |
| Hyperculture | 0.31 | 0.26 | 6.7% | <.01 |
| T&D in Scientific Knowledge and Methods | 0.11 | 0.09 | 0.8% | <.01 |
| Conscientiousness | -0.09 | -0.09 | 0.8% | <.01 |
| Age (Years) | -0.08 | -0.08 | 0.6% | <.01 |
| T&D inGgeneral | 0.09 | 0.07 | 0.5% | <.01 |
| Neuroticism | -0.04 | -0.04 | 0.2% | <.01 |
| IQ | 0.05 | 0.04 | 0.2% | <.01 |
| Being Male | 0.04 | 0.04 | 0.1% | <.01 |

The final model was statistically significant, accounting for a substantial proportion of variance in Sophotechnia (around 55%). Experience with GenAIs in months emerged as the strongest predictor, uniquely explaining 17.0% of the variance, followed by Hyperculture, which accounted for an additional 6.7 percent of unique variance. Smaller but statistically significant positive contributions were observed for training and development in scientific knowledge and methods, general training and development, and male sex. Conscientiousness, age, and Neuroticism showed small negative associations with Sophotechnia. All retained predictors were statistically significant at p<.01, indicating that Sophotechnic Mediation is primarily driven by direct experience with GenAIs and hypercultural engagement, with more modest contributions from cognitive, educational, personality, and demographic factors.

**The Development of Sophotechnia and the Role of Age**

Figure 7 shows a plot of means and confidence intervals for Sophotechnia as a function of different levels of experience with GenAIs.

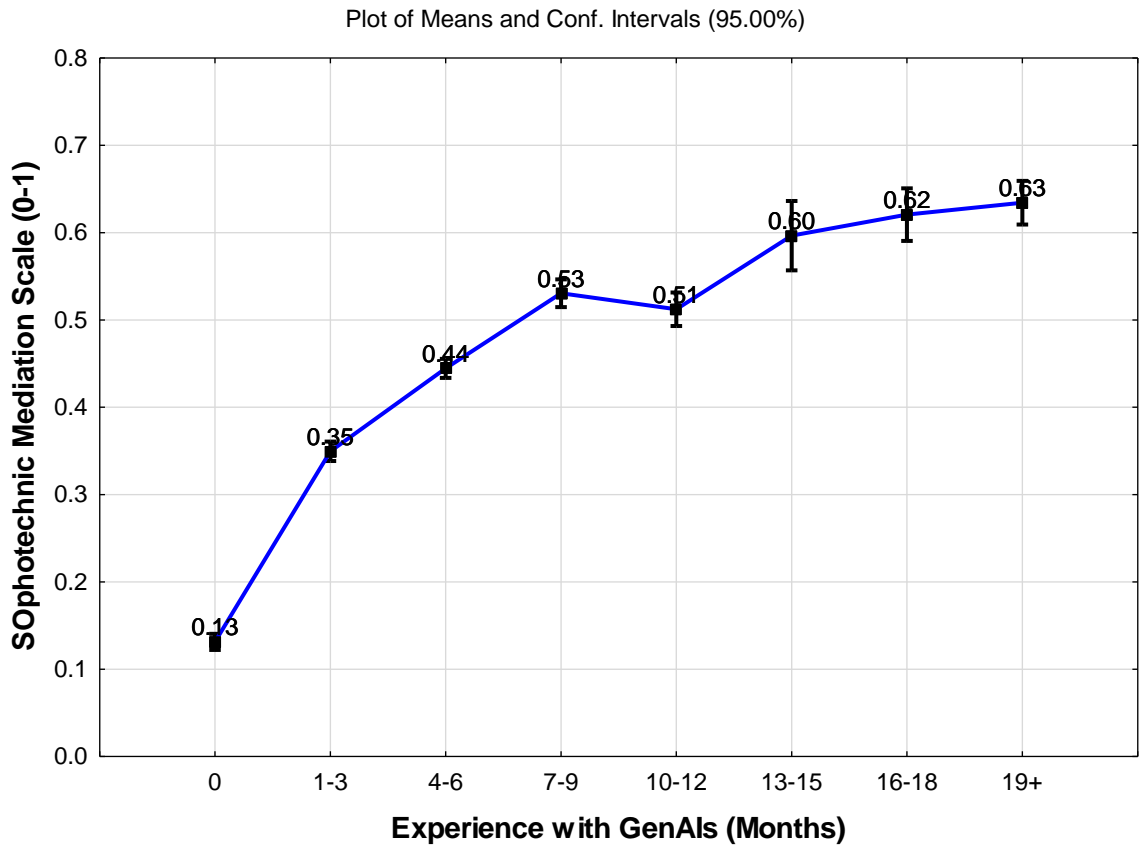

Figure 7: Sophotechnic Mediation Scale scores versus experience with GenAI.

There seems to be a clear non-linear pattern where an increase in the amount of experience with GenAIs is associated with an increase in Sophotechnia, with asymptotic diminishing returns, but also a significant "break" at around 10-12 months, after which there is a "jump" followed by a slow growth.

Figure 8 presents the plot of means and confidence intervals for Sophotechnia as a function of different levels of experience with GenAIs, subdivided according to the age in which the individual begins interacting with the technology (up to 25 years, 26 to 35 years, and 36 years or more).

Sophotechnia increased systematically with months of GenAI use in all age groups, indicating a strong cumulative experience effect. At low levels of experience, participants who began using GenAIs up to age 25 showed higher initial scores. However, after approximately 12 months of experience, participants who began between ages 26 and 35 and those who began at age 36 or later exhibited higher mean Sophotechnia scores than the early-onset group. This late-emerging advantage of older onset groups persisted at higher experience levels, with partially non-overlapping confidence intervals in the upper experience ranges. These results suggest that early exposure is associated with higher initial Sophotechnic Mediation scores at low experience levels, whereas later age at onset is associated with higher conditional mean scores at higher levels of accumulated experience.

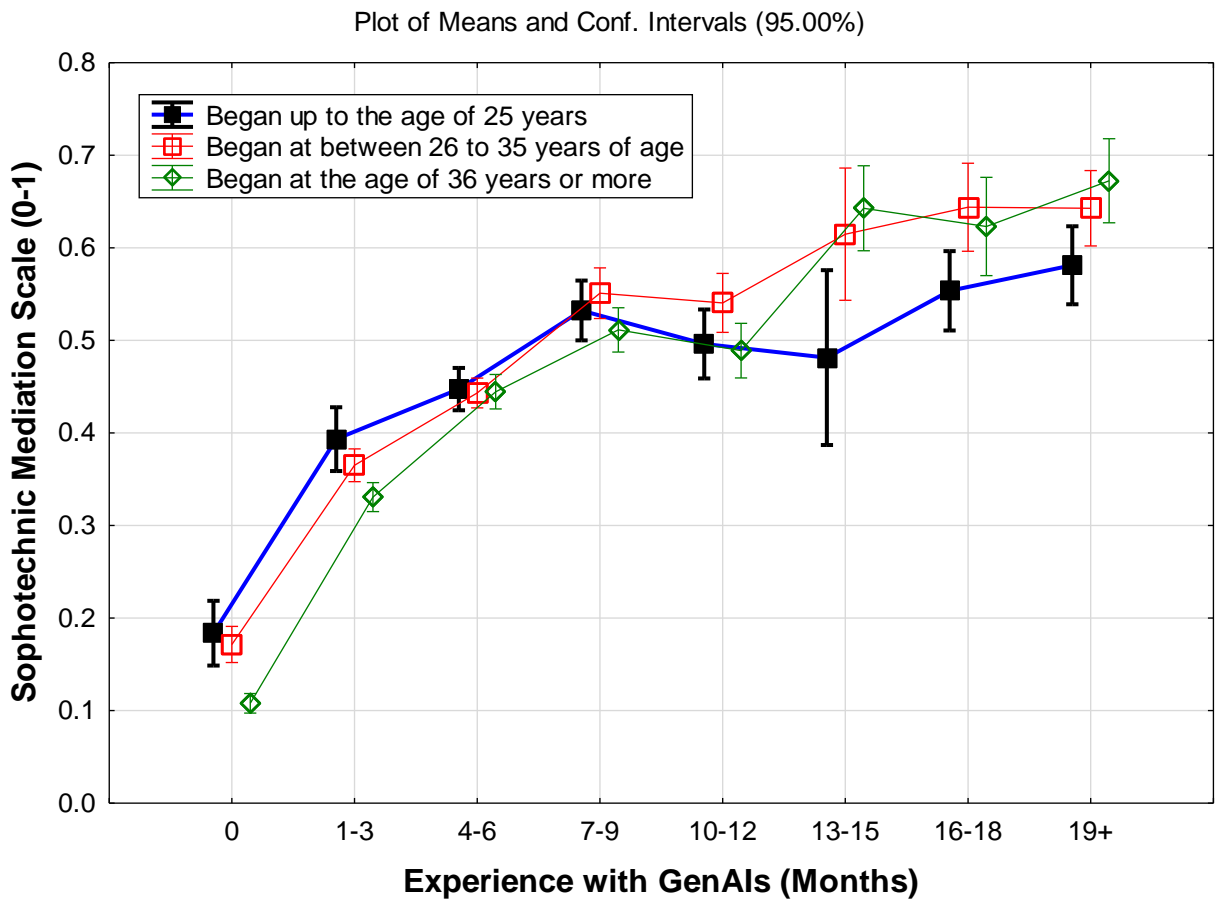

Figure 8: Sophotechnic Mediation Scale scores versus experience with GenAI for three age groups.

The association between the experience with GenAIs and the individual items of the Sophotechnic Mediation Scale revealed a pattern that was similar to that of the general score with regard to the variations in the pattern between the three age groups considered. However, this occurred only in three specific elements: "Experience with GenAI Groups, Communities and Forums", "Use of GenAI Browser Extensions", and "Frequency of Consuming News on GenAIs".

Figure 9 shows the plot of means and confidence intervals for the Experience with GenAI Groups, Communities and Forums as a function of different levels of experience with GenAIs, subdivided by age.

For those with age up to 25 years when they began interacting with GenAIs, the Experience with GenAI Groups, Communities and Forums increased with the amount of experience up to six months, after which it remains fairly constant. Among those who were between the ages of 26 and 35, the level of engagement rose constantly with the amount of experien ce, surpassing the younger group after six months. The group aged 36 years or more was behind the other, younger groups, at the start of its experience, but caught up with them in the first six months, reached the level of the 26–35-year-olds between seven and 12 months, and surpassed them after one year.

Figure 10 shows the plot of means and confidence intervals for the Frequency of Consuming News on GenAIs as a function of different levels of experience with GenAIs, subdivided by age.

The behavior of the Frequency of Consuming N ews on GenAIs as a function of experience with the technology, subdivided by age group, followed a pattern similar to that of the Experience with GenAI Groups, Communities and Forums, except that the youngest group reaches level close to that of those aged 26-35 years after 18 months.

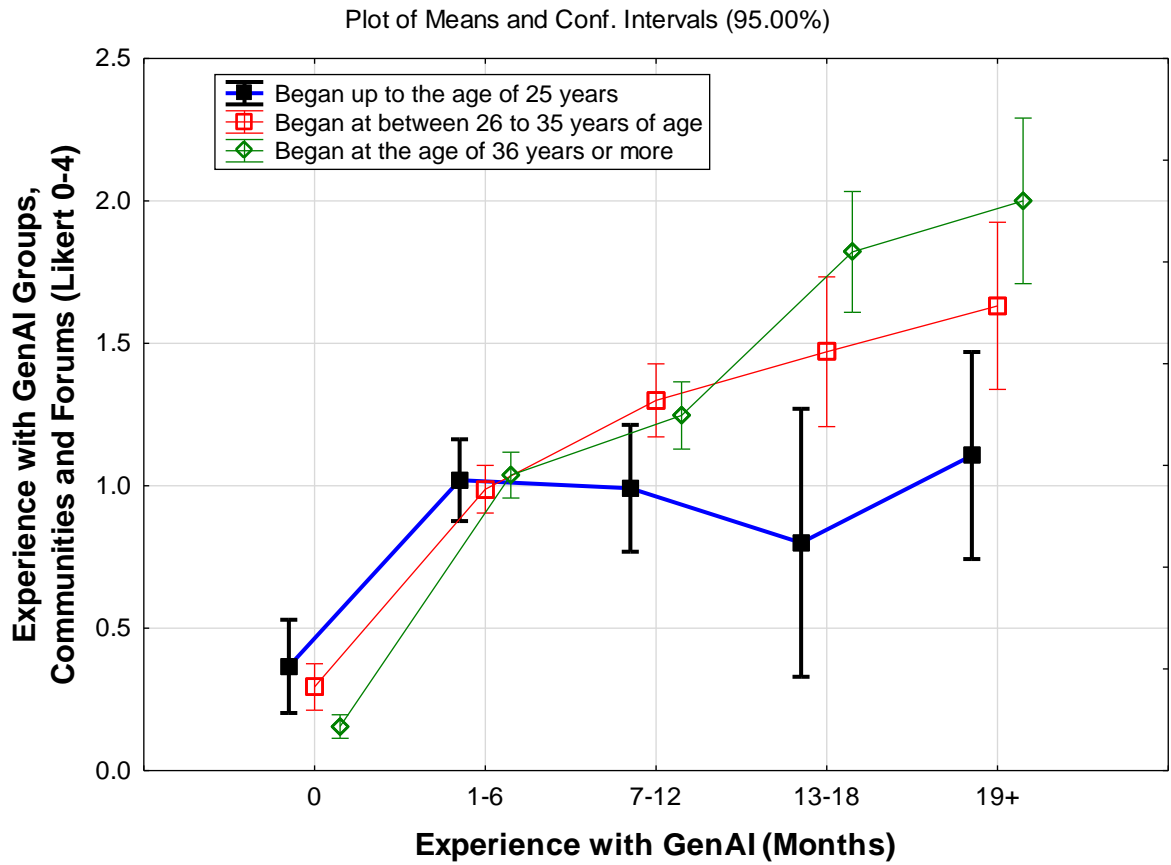

Figure 9: Engagement with GenAI groups versus experience with GenAI for three age groups.

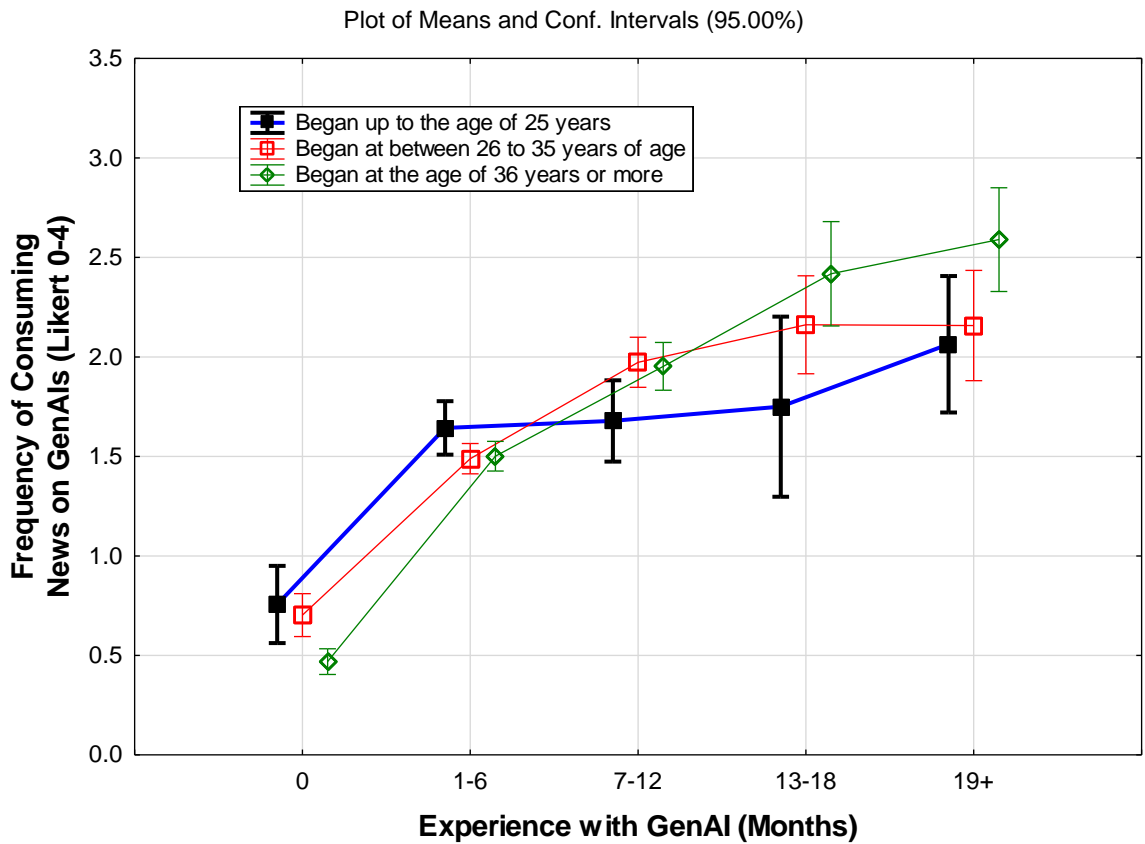

Figure 10: Engagement with GenAI Groups versus experience with GenAI for three age groups.

Figure 11 shows the plot of means and confidence intervals for the Use of GenAI Browser Extensions as a function of different levels of experience with GenAIs, subdivided by age.

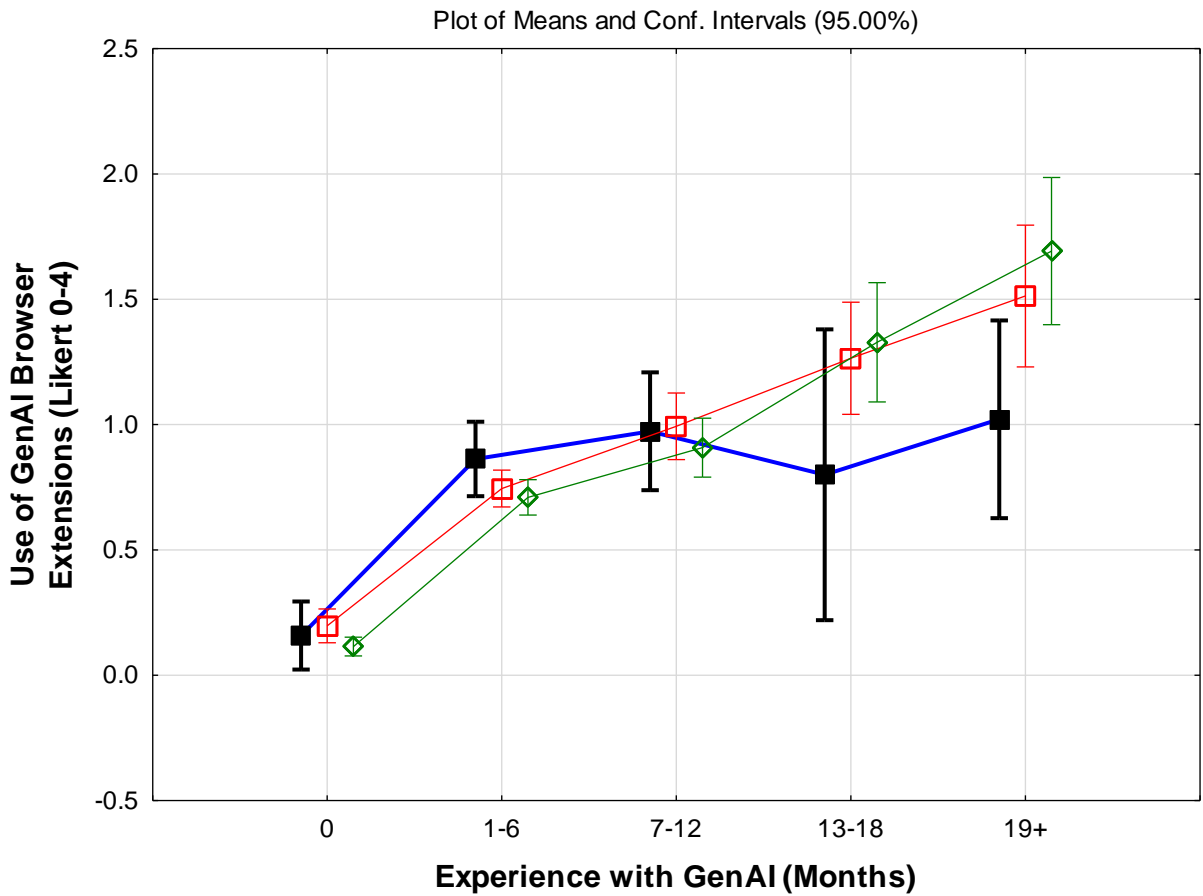

Figure 11: Use of GenAI Extensions versus experience with GenAI for three age groups.

The behavior of the Frequency of Using GenAI Browser Extensions as a function of experience with the technology, subdivided by age group, followed a pattern similar to that of the Experience with GenAI Groups, Communities and Forums.

Figure 12 shows the plot of means and confidence intervals for Sophotechnia as a function of different levels of experience with GenAIs, subdivided by levels of Hyperculture.

In both hypercultural groups, Sophotechnia increased consistently with months of GenAI experience, indicating a strong cumulative exposure effect. However, across all experience levels, individuals with a Hypercultural Index of .60 or higher exhibited higher Sophotechnic Mediation scores than those below .60. This advantage was already evident at the earliest stages of GenAI exposure and remained stable as experience increased. The gap between hypercultural groups did not diminish with additional experience, suggesting an additive rather than compensatory effect of GenAI use on Sophotechnia. Notably, at higher experience levels, the high-hyperculture group approached higher asymptotic Sophotechnia values, whereas the low-hyperculture group showed a slower increase and greater variability, particularly in the mid-range of experience. Overall, these findings indicate that hypercultural engagement systematically amplifies the development of Sophotechnia across the entire range of GenAI experience.

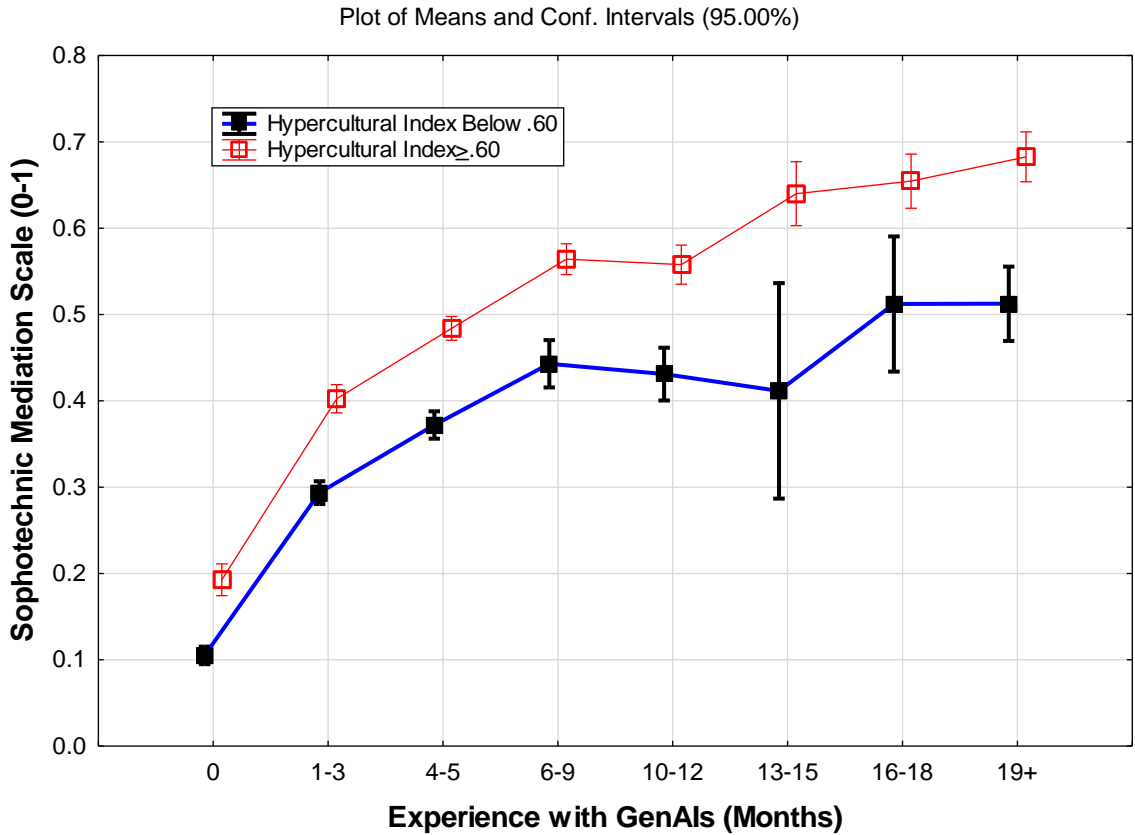

Figure 12: Sophotechnic Mediation versus experience with GenAI for two levels of Hyperculture.

## DISCUSSION

### Overview

The present study provides a comprehensive psychometric validation of the Sophotechnic Mediation Scale. It also examines the structural and theoretical properties of Sophotechnia as an emergent mode of cognitive mediation associated with sustained interaction with generative AI systems. Taken together, these results indicate that Sophotechnia constitutes a coherent and measurable construct whose expression is systematically structured by experience, age-related context, and sociotechnical engagement, consistent with the predictions of the Cognitive Mediation Networks Theory (CMNT), and empirically distinguishable from earlier forms of digital mediation such as Hyperculture.

### Psychometric Properties and Construct Coherence

The Sophotechnic Mediation Scale exhibited excellent internal consistency, with high Cronbach's alpha and item–total correlations, and no indication that the removal of any item would improve reliability. These results suggest that the scale items capture a tightly integrated construct rather than loosely related facets. The unidimensional factor structure further supports this conclusion, indicating that Sophotechnia is best conceptualized as a single latent dimension rather than a multidimensional skill set.

From a theoretical standpoint, this unidimensionality aligns with CMNT's conception of mediation as an integrated mode of functioning that emerges from the internalization of stable operational regularities of external systems. Rather than reflecting separable competencies such as frequency of use, technical knowledge, or ethical awareness in isolation, Sophotechnia appears to represent an internally coherent pattern of thinking and acting that integrates these elements into a unified mediation regime.

**Distributional Dynamics and Emergence of Sophotechnia**

The temporal evolution of the Sophotechnic Mediation Score provides strong evidence that Sophotechnia is an emergent mediation construct rather than a static individual difference. Across the three observation years, mean scores increased systematically, while dispersion among adopters remained relatively stable. At the same time, the proportion of zero scores declined sharply, indicating a rapid reduction in non-adoption.

Formal modeling rejected the interpretation that the observed distributions represent a single Gaussian process with a censored lower tail. The failure of the Tobit model and the clear superiority of the hurdle model indicate that adoption and intensity are governed by distinct processes. In early phases, the distribution reflects a mixture of non-adopters and adopters, whereas in later phases the non-adopter mass collapses and the conditional distribution among adopters converges toward approximate Gaussianity.

This pattern is consistent with diffusion-of-innovation dynamics as an interpretive framework, whose population-level distribution evolves over time as adoption spreads and experience accumulates. Importantly, the results indicate that early non-Gaussianity should not be treated as a psychometric anomaly, but as an expected feature of an emergent mediation regime.

**Differentiation Between Hyperculture and Sophotechnia**

Multiple analyses converged on the conclusion that Hypercultural mediation and Sophotechnic mediation are empirically distinguishable constructs. Cluster analysis and Smallest Space Analysis both revealed a clear structural separation between items of the two instruments, consistent with an axial partition. Although the two constructs are correlated, they form distinct clusters and cannot be reduced to a single dimension of digital engagement.

The non-linear association observed between Hyperculture and Sophotechnia further clarifies their relationship. At low levels of Hyperculture, the association with Sophotechnia is weak, suggesting that basic digital familiarity is not sufficient for the development of Sophotechnic Mediation. In an intermediate range, the association becomes stronger, indicating that Hyperculture provides a necessary but not sufficient foundation for Sophotechnia. At higher levels, the association attenuates again, implying that beyond a certain point, Sophotechnia depends primarily on direct interaction with GenAI systems rather than on general digital sophistication.

These findings are consistent with CMNT's layered view of mediation, according to which new mediation regimes build upon, but do not collapse into, earlier ones.

**Experience With GenAI as the Primary Driver of Development**

Across all analyses, experience with GenAI systems emerged as the dominant predictor of Sophotechnia, uniquely accounting for a substantial proportion of variance. The observed growth curve was markedly non-linear, with diminishing returns at higher levels of experience but a clear qualitative shift around 10 to 12 months of sustained use.

This breakpoint is theoretically meaningful. It likely reflects a transition from exploratory or instrumental use toward a more internalized mode of interaction, characterized by improved orchestration, strategic prompting, metacognitive monitoring, and integration of GenAI into complex cognitive workflows.

Within CMNT, this transition corresponds to the internalization of functional regularities of the external system, resulting in a stable mediation regime rather than simply an episodic tool use.

**Developmental Windows and Age-Related Effects**

When Sophotechnia is treated as a developmental mode of cognitive mediation rather than a simple accumulation of technical competencies, it becomes necessary to distinguish developmental windows that differentially affect the rate of acquisition and the depth of integration of its constituent components. In this context, the age ranges of up to 25 years, 26–35 years, and 36 years or more delineate qualitatively distinct regimes through which Sophotechnic Mediation emerges, reorganizes, and reaches asymptotic expression.

The age range up to approximately 25 years corresponds to a late developmental integration window marked by ongoing maturation of prefrontal and frontoparietal control networks. Neurodevelopmental processes during this period support rapid learning, exploratory behavior, and flexible adaptation to novel cognitive tools. Consistent with this profile, individuals who begin interacting with generative artificial intelligence systems earlier tend to exhibit higher initial levels of Sophotechnic Mediation at low experience levels. This early advantage is evident in the global Sophotechnic score and reflects faster acquisition of basic interaction patterns rather than long-term optimization.

However, component-level analyses reveal that this early advantage is transient. As shown in the figures depicting engagement in GenAI communities and forums, use of ChatGPT browser extensions, and frequency of consuming news about GenAIs, younger starters do not maintain their lead as experience accumulates. Instead, after approximately 7–12 months of sustained GenAI use, individuals who began using these systems later in life consistently overtake the younger group across all three components. This crossover pattern indicates that an earlier age at onset is associated with faster initial uptake of GenAI-related practices, whereas deeper sociotechnical integration appears more strongly associated with accumulated professional experience and contextual factors.

The 26–35 age range represents a post-maturational regime in which executive and control networks have largely stabilized. At this stage, Sophotechnic development increasingly reflects strategic learning, accumulated domain knowledge, and the ability to integrate GenAI into complex professional and cognitive workflows. Empirically, individuals who began using GenAI during this period show steeper mid- to late-stage growth in component-level indicators, particularly in the adoption of browser extensions and sustained engagement with GenAI-related information ecosystems. These patterns suggest that stabilized cognitive architectures combined with growing experience favor the internalization of higher-order mediation practices.

The 36 years or more range corresponds to a regime of consolidation and optimization rather than ongoing maturation. In this group, later starters consistently achieve the highest asymptotic levels in components that index ecosystem-level Sophotechnia. Participation in GenAI communities, systematic use of infrastructural extensions, and regular consumption of specialized news about GenAI all show stronger late-stage growth and higher terminal means among older participants once sufficient experience is accumulated. These findings indicate that accumulated professional expertise, institutional embeddedness, and metacognitive control enable deeper integration of GenAI into stable cognitive and organizational routines, even when initial learning is slower.

Crucially, the convergence of these component-level patterns across multiple indicators demonstrates that the age-related crossover is not an artifact of a single behavior or item, but reflects a structural property of Sophotechnic development. Early exposure primarily accelerates the entry into Sophotechnic Mediation, whereas later exposure supports more advanced orchestration of GenAI within broader sociotechnical systems. This distinction clarifies why older starters ultimately surpass younger ones in behaviors that require strategic selectivity, sustained monitoring of the field, and participation in communities of practice.

Taken together, these results indicate that age-related context moderates Sophotechnia along two separable dimensions. Age influences the rate of early acquisition, while accumulated life and professional experience are associated with higher conditional levels of mediation at greater experience, particularly at the ecosystem level. Age should therefore not be interpreted as a linear facilitator or constraint on Sophotechnic Mediation. Instead, it functions as a moderator that determines how experience with generative artificial intelligence is transformed into stable, internalized modes of cognitive mediation. This interpretation strengthens the Cognitive Mediation Networks Theory by showing that Sophotechnia emerges from the interaction between sustained GenAI experience and the individual's pre-existing cognitive, professional, and sociotechnical structures, rather than from neurobiological plasticity alone.

**Hyperculture as an Amplifying Factor**

Hypercultural mediation consistently amplified Sophotechnia across all levels of GenAI experience. Individuals with higher Hypercultural Index scores exhibited higher Sophotechnic Mediation scores from the earliest stages of adoption, and this advantage remained stable as experience increased. The persistence of this gap indicates an additive rather than compensatory relationship between Hyperculture and Sophotechnia.

This finding reinforces the view that Sophotechnia emerges within a broader sociotechnical ecology. Hypercultural engagement appears to facilitate access to practices, norms, and informational resources that support the internalization of Sophotechnic Mediation, without substituting for direct interaction with GenAI systems.

**Secondary Predictors and Individual Differences**

In addition to experience and Hyperculture, several secondary predictors contributed modest but statistically significant effects. Training in scientific knowledge and methods, general training and development, IQ, and male sex were positively associated with Sophotechnia, whereas Conscientiousness, Neuroticism, and age showed small negative associations.

These effects are theoretically interpretable. Higher cognitive ability and formal training likely facilitate abstraction and transfer, supporting more efficient internalization of AI-mediated strategies. The negative association with Conscientiousness may reflect reduced experimentation or risk-taking in AI use, while higher Neuroticism may interfere with exploratory engagement. Importantly, the relatively small effect sizes of these variables underscore that Sophotechnia is not primarily determined by stable personality or cognitive traits, but by sustained engagement with GenAI systems within supportive sociotechnical contexts.

**Implications for Cognitive Mediation Networks Theory**

The findings provide strong empirical support for CMNT's central claims. Sophotechnia behaves as a distinct mediation regime that emerges through repeated interaction with a novel class of external mechanisms and becomes internalized over time. Unlike accounts focused solely on cognitive offloading or tool dependence, the present results indicate that, consistent with an internalization process, GenAI engagement is associated with stable, measurable differences.

Moreover, the observed cohort-wise experience patterns distinguish Sophotechnia from earlier forms of digital mediation, supporting the claim that generative AI represents a qualitative technological discontinuity rather than a mere extension of existing tools.

**Practical Implications**

From an applied perspective, the results suggest that access to GenAI systems alone is insufficient to foster Sophotechnia. Meaningful development depends on sustained use, integration into complex tasks, and embedding within hypercultural environments that support learning and experimentation. These findings have direct implications for organizational training, workforce development, and educational policy, particularly in contexts where unequal access to GenAI may translate into long-term cognitive and professional disparities.

**Limitations and Future Directions**

Several limitations warrant consideration. The study relied on cross-sectional data from independent yearly samples, precluding direct within-individual developmental inference. The sample was geographically restricted, and all measures were self-reported. Future research should incorporate longitudinal designs, behavioral and performance-based validation, and cross-cultural replication to further test the robustness and generalizability of the construct.

All age-related findings should be interpreted as cross-sectional moderation effects rather than as direct evidence of intra-individual developmental change. Although the theoretical framework draws on developmental neuroscience, the present design does not permit causal inferences about neurodevelopmental mechanisms. Longitudinal and experimental studies will be required to directly test developmental trajectories of Sophotechnic Mediation

**Conclusion**

In conclusion, the present study provides strong evidence that Sophotechnia is a coherent and measurable mode of cognitive mediation whose expression is systematically structured by cumulative GenAI experience and age-related context. The Sophotechnic Mediation Scale demonstrates robust psychometric properties and theoretical validity, positioning it as a foundational instrument for future research on cognition and work in the era of generative AI.

# APPENDIX

# Original Form in Brazilian Portuguese

### QUESTIONÁRIO DE SOFOTECNIA

As seguintes perguntas se referem às suas interações e experiências com ferramentas de Inteligência Artificial (IA), como o ChatGPT, Microsoft CoPilot ou Google Gemini, que entendem suas perguntas em linguagem cotidiana e processam informações complexas para fornecer respostas detalhadas.

01) Você usa ou já usou alguma ferramenta de IA desse tipo?

(0) Nunca
(1) Apenas uma ou duas vezes
(2) Algumas vezes
(3) Frequentemente
(4) O tempo todo

02) Quais das ferramentas de IA generativa abaixo você usa ou já usou?

A) Chatbots, Large Language Models (LLM) e afins:

| | | | |
|---|---|---|---|
| i) OpenAI ChatGPT | (0) Não | (1) Sim, versão gratuita | (2) Sim, versão paga |
| ii) Microsoft CoPilot (Antigo Bing Chat) | (0) Não | (1) Sim, versão gratuita | (2) Sim, versão paga |
| iii) Google Gemini (Antigo Google Bard) | (0) Não | (1) Sim, versão gratuita | (2) Sim, versão paga |
| iv) Anthropic Claude | (0) Não | (1) Sim, versão gratuita | (2) Sim, versão paga |
| v) X Grok | (0) Não | (1) Sim, versão gratuita | (2) Sim, versão paga |
| vi) DeepSeek | (0) Não | (1) Sim, versão gratuita | (2) Sim, versão paga |
| vii) Perplexity | (0) Não | (1) Sim, versão gratuita | (2) Sim, versão paga |
| viii) Outras | (0) Não | (1) Sim, versão gratuita | (2) Sim, versão paga |

B) Sistemas voltados especificamente para a produção de:

| | | | |
|---|---|---|---|
| i) Imagens | (0) Não | (1) Sim, versão gratuita | (2) Sim, versão paga |
| ii) Vídeos | (0) Não | (1) Sim, versão gratuita | (2) Sim, versão paga |
| iii) Música | (0) Não | (1) Sim, versão gratuita | (2) Sim, versão paga |
| iv) Apresentações | (0) Não | (1) Sim, versão gratuita | (2) Sim, versão paga |
| v) Podcasts | (0) Não | (1) Sim, versão gratuita | (2) Sim, versão paga |
| vi) Software | (0) Não | (1) Sim, versão gratuita | (2) Sim, versão paga |
| vii) Outros | (0) Não | (1) Sim, versão gratuita | (2) Sim, versão paga |

03) Você realiza ou já realizou alguma das seguintes tarefas com a ajuda de alguma ferramenta de IA?

| | | |
|---|---|---|
| A) Resumir e/ou interpretar um texto. | (1) Sim | (0) Não |
| B) Escrever um ensaio, relatório, resenha, artigo ou outro texto. | (1) Sim | (0) Não |
| C) Organizar tarefas ou horários. | (1) Sim | (0) Não |
| D) Aprender sobre algum assunto. | (1) Sim | (0) Não |
| E) Gerar ideias ou sugestões. | (1) Sim | (0) Não |
| F) Pesquisar na internet e/ou organizar resultados de busca. | (1) Sim | (0) Não |
| G) Analisar ou interpretar dados qualitativos. | (1) Sim | (0) Não |
| H) Analisar ou interpretar dados quantitativos. | (1) Sim | (0) Não |

I) Explorar relações entre conceitos, teorias e ideias.  (1) Sim   (0) Não
J) Codificação ou programação.  (1) Sim   (0) Não
K) Fazer apresentações  (1) Sim   (0) Não
l) Produzir imagens, vídeos, música ou outros conteúdos multimídia.  (1) Sim   (0) Não

04) Usando a escala abaixo, avalie o quanto você confia na sua habilidade de fazer perguntas específicas de forma a produzir a resposta ou comportamento desejado ao usar alguma ferramenta de IA.

(0) Nada confiante
(1) Pouco confiante
(2) Algo confiante
(3) Muito confiante
(4) Extremamente confiante

05) Você está ciente das limitações e dos possíveis vieses (tendenciosidades) nas informações fornecidas por tais ferramentas de IA?

(0) Não estou ciente   (1) De poucos   (2) De alguns   (3) De muitos   (4) Da maioria ou de todos

06) Usando a escala abaixo, avalie seu entendimento sobre a diferença entre as informações geradas por ferramentas digitais com inteligência artificial e as pesquisas gerais na internet.

(0) Eu não sabia que havia diferença
(1) Eu sei que há diferença, mas não sei qual é
(2) Tenho alguma ideia da diferença
(3) Conheço bem a diferença
(4) Tenho um entendimento profundo da diferença

07)     Há quantos meses você interage regularmente com ferramentas digitais movidas a IA para processamento de informações, além de pesquisas básicas na internet e ferramentas de comunicação?

_______ meses (colocar "0" se há menos de um mês)

08)     Usando a escala abaixo, avalie sua experiência com grupos, comunidades ou fóruns de discussão focados em ferramentas digitais movidas a IA e tecnologias relacionadas, em oposição ao uso geral da internet e da tecnologia.

(0) Nenhuma experiência
(1) Pouca experiência
(2) Alguma experiência
(3) Experiência substancial
(4) Muita experiência

09) Usando a escala abaixo, avalie o quanto você confia na sua capacidade de reconhecer a diferença entre conteúdo gerado por IA e conteúdo produzido por meio de pesquisas gerais na internet.

(0) Nada confiante
(1) Pouca confiança
(2) Mais ou menos confiante
(3) Confiante
(4) Extremamente confiante

10) Usando a escala abaixo, com que frequência você conta com ferramentas digitais movidas a IA para ajudá-lo a tomar decisões informadas ou resolver problemas complexos, em oposição a apenas pesquisas básicas na Internet e ferramentas de comunicação?

(0) Nunca    (1) Raramente    (2) Às vezes    (3) Frequentemente    (4) O tempo todo

11) Usando a escala abaixo, avalie seu grau de conhecimento das diretrizes ou considerações éticas especificamente relacionadas ao uso de ferramentas digitais movidas a IA para processamento de informações, além do uso geral da internet e das ferramentas de comunicação.

(0) Nenhum    (1) Um pouco    (2) Algum    (3) Grande    (4) Extenso

12) Você usa ou já usou extensões de navegador ou complementos semelhantes para expandir a funcionalidade do ChatGPT?

(0) Nunca    (1) Raramente    (2) Às vezes    (3) Frequentemente    (4) O tempo todo

13) Com que frequência você interage com ferramentas digitais movidas a IA por meio de uma conversa para explorar um determinado tópico e expandir seu entendimento sobre ele?

(0) Nunca    (1) Raramente    (2) Às vezes    (3) Frequentemente    (4) O tempo todo

14) Com que frequência você acompanha as notícias, os últimos desenvolvimentos e eventuais atualizações em ferramentas de IA e tecnologias relacionadas?

(0) Nunca    (1) Raramente    (2) Às vezes    (3) Frequentemente    (4) O tempo todo

15) O seu uso e entendimento de ferramentas de IA como ChatGPT, CoPilot, Gemini ou tecnologias semelhantes afetaram a maneira como você pensa sobre o que é o conhecimento, sobre como se aprende e/ou sobre como resolver problemas?

(0) De forma alguma   (1) Um pouco    (2) Em alguma medida   (3) Muito   (4) Profundamente

16) A sua empresa tem alguma iniciativa usando ferramentas de IA como ChatGPT, CoPilot, Gemini ou tecnologias semelhantes nas suas operações?

(0) Não, nem se fala no assunto.
(1) Não, mas se está considerando seriamente a possibilidade.
(2) Existem planos concretos sendo feitos, mas ainda sem nada iniciado na prática.
(3) Sim, mas ainda em fase de implementação.
(4) Sim, já implementado e em funcionamento.
(5) Sim, já funcionando e se considerando seriamente expandir.
(6) Sim, já funcionando e com expansão em andamento.

17) Na sua opinião, que tipo de impacto o uso de ferramentas de IA como ChatGPT, CoPilot, Gemini ou tecnologias semelhantes na sua empresa tem ou terá:

A) Para a sua empresa?                            (0) Negativo   (1) Neutro    (2) Positivo
B) Para você?                                     (0) Negativo   (1) Neutro    (2) Positivo
C) Para o mercado de trabalho da sua ocupação? (0) Negativo   (1) Neutro     (2) Positivo

# Translation Into American English

**SOPHOTECHNICS QUESTIONNAIRE**

The following questions refer to your interactions and experiences with Artificial Intelligence (AI) tools such as ChatGPT, Microsoft Copilot, or Google Gemini, which understand questions in everyday language and process complex information to provide detailed answers.

01) Do you use or have you ever used any AI tool of this kind?

(0) Never
(1) Only once or twice
(2) A few times
(3) Frequently
(4) All the time

02) Which of the following generative AI tools do you use or have you used?

A) Chatbots, Large Language Models (LLMs), and similar tools:

| | | | |
|---|---|---|---|
| i) OpenAI ChatGPT | (0) No | (1) Yes, free version | (2) Yes, paid version |
| ii) Microsoft Copilot (formerly Bing Chat) | (0) No | (1) Yes, free version | (2) Yes, paid version |
| iii) Google Gemini (formerly Google Bard) | (0) No | (1) Yes, free version | (2) Yes, paid version |
| iv) Anthropic Claude | (0) No | (1) Yes, free version | (2) Yes, paid version |
| v) X Grok | (0) No | (1) Yes, free version | (2) Yes, paid version |
| vi) DeepSeek | (0) No | (1) Yes, free version | (2) Yes, paid version |
| vii) Perplexity | (0) No | (1) Yes, free version | (2) Yes, paid version |
| viii) Others | (0) No | (1) Yes, free version | (2) Yes, paid version |

B) Systems specifically aimed at producing:

| | | | |
|---|---|---|---|
| i) Images | (0) No | (1) Yes, free version | (2) Yes, paid version |
| ii) Videos | (0) No | (1) Yes, free version | (2) Yes, paid version |
| iii) Music | (0) No | (1) Yes, free version | (2) Yes, paid version |
| iv) Presentations | (0) No | (1) Yes, free version | (2) Yes, paid version |
| v) Podcasts | (0) No | (1) Yes, free version | (2) Yes, paid version |
| vi) Software | (0) No | (1) Yes, free version | (2) Yes, paid version |
| vii) Others | (0) No | (1) Yes, free version | (2) Yes, paid version |

03) Do you perform or have you ever performed any of the following tasks with the help of an AI tool?

| | | |
|---|---|---|
| A) Summarizing and or interpreting a text. | (1) Yes | (0) No |
| B) Writing an essay, report, review, article, or other text. | (1) Yes | (0) No |
| C) Organizing tasks or schedules. | (1) Yes | (0) No |
| D) Learning about a topic. | (1) Yes | (0) No |
| E) Generating ideas or suggestions. | (1) Yes | (0) No |
| F) Searching the internet and or organizing search results. | (1) Yes | (0) No |
| G) Analyzing or interpreting qualitative data. | (1) Yes | (0) No |
| H) Analyzing or interpreting quantitative data. | (1) Yes | (0) No |
| I) Exploring relationships among concepts, theories, and ideas. | (1) Yes | (0) No |
| J) Coding or programming. | (1) Yes | (0) No |

K) Creating presentations. (1) Yes (0) No
L) Producing images, videos, music, or other multimedia content. (1) Yes (0) No

04) Using the scale below, rate how confident you are in your ability to ask specific questions in order to produce the desired response or behavior when using an AI tool.

(0) Not confident at all
(1) Slightly confident
(2) Somewhat confident
(3) Very confident
(4) Extremely confident

05) Are you aware of the limitations and possible biases in the information provided by such AI tools?

(0) Not aware
(1) Aware of a few
(2) Aware of some
(3) Aware of many
(4) Aware of most or all

06) Using the scale below, rate your understanding of the difference between information generated by AI-powered digital tools and general internet searches.

(0) I did not know there was a difference
(1) I know there is a difference, but I do not know what it is
(2) I have some idea of the difference
(3) I know the difference well
(4) I have a deep understanding of the difference

07) For how many months have you been regularly interacting with AI-powered digital tools for information processing, beyond basic internet searches and communication tools?

\_\_\_\_\_\_ months (enter "0" if less than one month)

08) Using the scale below, rate your experience with groups, communities, or discussion forums focused on AI-powered digital tools and related technologies, as opposed to general internet and technology use.

(0) No experience
(1) Little experience
(2) Some experience
(3) Substantial experience
(4) Extensive experience

09) Using the scale below, rate how confident you are in your ability to recognize the difference between AI-generated content and content produced through general internet research.

(0) Not confident at all
(1) Slightly confident
(2) Moderately confident
(3) Confident
(4) Extremely confident

10) Using the scale below, how often do you rely on AI-powered digital tools to help you make informed decisions or solve complex problems, as opposed to only basic internet searches and communication tools?

(0) Never   (1) Rarely   (2) Sometimes   (3) Frequently   (4) All the time

11) Using the scale below, rate your level of knowledge of guidelines or ethical considerations specifically related to the use of AI-powered digital tools for information processing, beyond general internet and communication tool use.

(0) None   (1) A little   (2) Some   (3) A great deal   (4) Extensive

12) Do you use or have you ever used browser extensions or similar add-ons to expand the functionality of ChatGPT?

(0) Never   (1) Rarely   (2) Sometimes   (3) Frequently   (4) All the time

13) How often do you interact with AI-powered digital tools through conversation to explore a given topic and expand your understanding of it?

(0) Never   (1) Rarely   (2) Sometimes   (3) Frequently   (4) All the time

14) How often do you follow the news, latest developments, and updates related to AI tools and related technologies?

(0) Never   (1) Rarely   (2) Sometimes   (3) Frequently   (4) All the time

15) Has your use and understanding of AI tools such as ChatGPT, Copilot, Gemini, or similar technologies affected the way you think about what knowledge is, how learning occurs, and or how problems are solved?

(0) Not at all   (1) A little   (2) To some extent   (3) A great deal   (4) Profoundly

16) Does your company have any initiative using AI tools such as ChatGPT, Copilot, Gemini, or similar technologies in its operations?

(0) No, the topic is not even discussed.
(1) No, but the possibility is being seriously considered.
(2) Concrete plans exist, but nothing has yet been implemented.
(3) Yes, but still in the implementation phase.
(4) Yes, implemented and in operation.
(5) Yes, in operation and seriously considering expansion.
(6) Yes, in operation with expansion underway.

17) In your opinion, what kind of impact does or will the use of AI tools such as ChatGPT, Copilot, Gemini, or similar technologies have:

| | |
|---|---|
| A) For your company? | (0) Negative   (1) Neutral   (2) Positive |
| B) For you? | (0) Negative   (1) Neutral   (2) Positive |
| C) For the job market in your occupation? | (0) Negative   (1) Neutral   (2) Positive |